\documentclass[aps,prb,preprint,superscriptaddress]{revtex4}

\usepackage{amsmath}
\usepackage{epsfig,bm}

\begin{document}


\title{Collective Mode at Lifshitz Transition in Iron-Pnictide Superconductors}

\author{Jose P. Rodriguez}

\affiliation{Department of Physics and Astronomy,
California State University at Los Angeles,
Los Angeles, California 90032.}


\begin{abstract}
We obtain the exact low-energy spectrum of two mobile holes in a 
$t$-$J$ model
for an isolated layer
in an iron-pnictide superconductor.
The minimum $d_{xz}$ and $d_{yz}$ orbitals per iron atom are included,
with no hybridization between the two.
After tuning the Hund coupling to a putative
quantum critical point (QCP) that separates 
a commensurate spin-density wave 
from a hidden-order antiferromagnet at half filling,
we find an $s$-wave hole-pair groundstate and a $d$-wave hole-pair excited state.
Near the QCP,
both alternate in sign between hole Fermi surface pockets 
at the Brillouin zone center
and emergent electron Fermi surface pockets at momenta
that correspond to commensurate spin-density waves (cSDW).
The dependence of the energy splitting with increasing Hund coupling
yields evidence for a true QCP in the thermodynamic limit near the putative one,
at which the $s$-wave and $d$-wave Cooper pairs are degenerate.
A collective $s$-to-$d$-wave oscillation of the macroscopic superconductor 
that couples to orthorhombic shear strain is also identified.
Its resonant frequency is predicted to collapse to
zero at the QCP in the limit of low hole concentration.
This implies degeneracy of
Cooper pairs with
$s$, $d$ and $s+id$ symmetry
in the corresponding quantum critical state.
We argue that the critical state describes
Cooper pairs in hole-doped iron superconductors 
at the Lifshitz transition, where electron bands first rise above the Fermi level.
We thereby predict that the $s$-to-$d$-wave collective mode observed by
Raman spectroscopy in Ba$_{1-x}$K$_x$Fe$_2$As$_2$ at optimal doping
should also be observed at higher doping near the Lifshitz transition.

\end{abstract}

\maketitle

\section{Introduction}
The symmetry of the Cooper pairs of electrons in iron-pnictide high-temperature superconductors
remains a subject of controversy. 
A stack of weakly coupled square lattices of iron atoms
is the common structural feature in these materials.
Early calculations of the electronic band structure within the density functional approximation
predicted nested two-dimensional (2D) Fermi surfaces\cite{singh_du_08,dong_08},
with hole pockets centered at zero 2D momentum,
and with electron pockets centered 
at 2D momenta $\hbar (\pi/a){\bm{\hat x}}$ and $\hbar (\pi/a){\bm{\hat y}}$.
Here $a$ is the intra-layer Fe-Fe separation.
The predicted Fermi surfaces where later confirmed experimentally 
by angle-resolved photo-emission spectroscopy (ARPES)\cite{ding_08,ding_09}.
Early calculations based on the above nested electronic structure
also predicted an $S^{+-}$ symmetry for the wavefunction of the Cooper pair 
that alternates in sign between the hole pockets and the electron pockets
because of spin-fluctuation exchange at 
the nesting vectors\cite{mazin_08,kuroki_08}.
Nearly degenerate Cooper pairs with $d$-wave symmetry are also predicted to exist
because of such spin-fluctuation exchange\cite{wang_09,graser_09}, however.

The discovery of the
hole-doped series Ba$_{1-x}$K$_x$Fe$_2$As$_2$
poses a challenge to the proposal of $S^{+-}$ Cooper pair symmetry
in iron-pnictide superconductors\cite{paglione_greene_10}.
High-temperature superconductivity exists
in a considerable range about optimal doping\cite{budko_prb_13}
at $x\cong 0.4$, where $T_c \cong 38$ K.
ARPES in this range of doping finds evidence for an $s$-wave gap
both on hole Fermi surfaces at zero 2D momentum
and on small electron Fermi surfaces at 
commensurate spin-density wave (cSDW) momenta\cite{ding_09,nakayama_prb_11}.
The low-temperature dependence of
the London penetration depth and of
the thermal conductivity at optimal doping\cite{hashimoto_prl_09,taillefer_prb_09}
are consistent with such $s$-wave superconducting gaps.
ARPES also finds that the small electron Fermi surface pockets
that exist at optimal doping
are {\it absent} in the end-member compound\cite{sato_prl_09} at $x = 1$,
however, where $T_c = 3$ K.
Specifically,
ARPES and thermo-electric measurements
find evidence for a continuous Lifshitz transition at
a concentration inside the range $0.5 < x < 1$,
where the electron bands at cSDW momenta
first rise above the Fermi level\cite{malaeb_prb_12,hodovanets_prb_14}.
A recent calculation based on density-functional theory (DFT) argues that the
Lifshitz transition is a result of the (Ba,K) alloy\cite{khan_prl_14}.
At optimal doping, however,
ARPES finds strong mass renormalizations of the electron bands at cSDW momenta
by up to a factor of six compared to DFT\cite{sato_prl_09}.
This failure suggests that strong on-site Coulomb repulsion
should be taken into account\cite{Si&A,jpr_ehr_09,jpr_10,jpr_mana_pds_11}.

We reveal the nature of a single Cooper pair in 
a local-moment $t$-$J$ model for hole-doped iron superconductors
characterized by large on-site Coulomb repulsion.
Two mobile holes roam over a $4\times 4$ periodic lattice
of spin-$1$ iron atoms that contain only $d_{xz}$ and $d_{yz}$ orbitals.
They coincide with the orbital character of the most intense bands seen 
by ARPES at the Brillouin zone center and at cSDW momenta 
in optimally doped\cite{malaeb_prb_12} Ba$_{1-x}$K$_x$Fe$_2$As$_2$.
The hole bands centered at zero 2D momentum are fixed by the band structure ($t$),
while the electron bands are emergent at nesting wave vectors
because of proximity to a cSDW state
that is favored by antiferromagnetic frustration ($J$)\cite{Si&A,jpr_ehr_09}.
In the case of one mobile hole,
both bands cross the Fermi level at a critical Hund coupling
that is of moderate strength\cite{jpr_mana_pds_14}.
A comparison with the results of
Schwinger-boson-slave-fermion meanfield theory suggests that the latter
coincides with a quantum critical point (QCP)
that separates a cSDW at strong Hund coupling
from a hidden-order magnet 
at weak Hund coupling \cite{jpr_mana_pds_14}.
Unlike previous exact numerical studies of pairing in two-orbital 2D superconductors
with on-site Coulomb repulsion\cite{dagotto_11},
we thereby avoid accounting for the electron pockets 
with fine-tuned hopping matrix elements\cite{raghu_08}
that result in an unphysical hole Fermi surface pocket\cite{graser_09}.
The theory necessarily predicts a Mott insulator state at half filling.
Recent experimental evidence for an insulator-superconductor transition
in single-layer FeSe supports this prediction\cite{zhou_14,sup_mat_2}.
It is related to the present hole-doped study 
by the application of an exact particle-hole transformation\cite{jpr_16b}.

\begin{figure}
\includegraphics[scale=0.70, angle=-90]{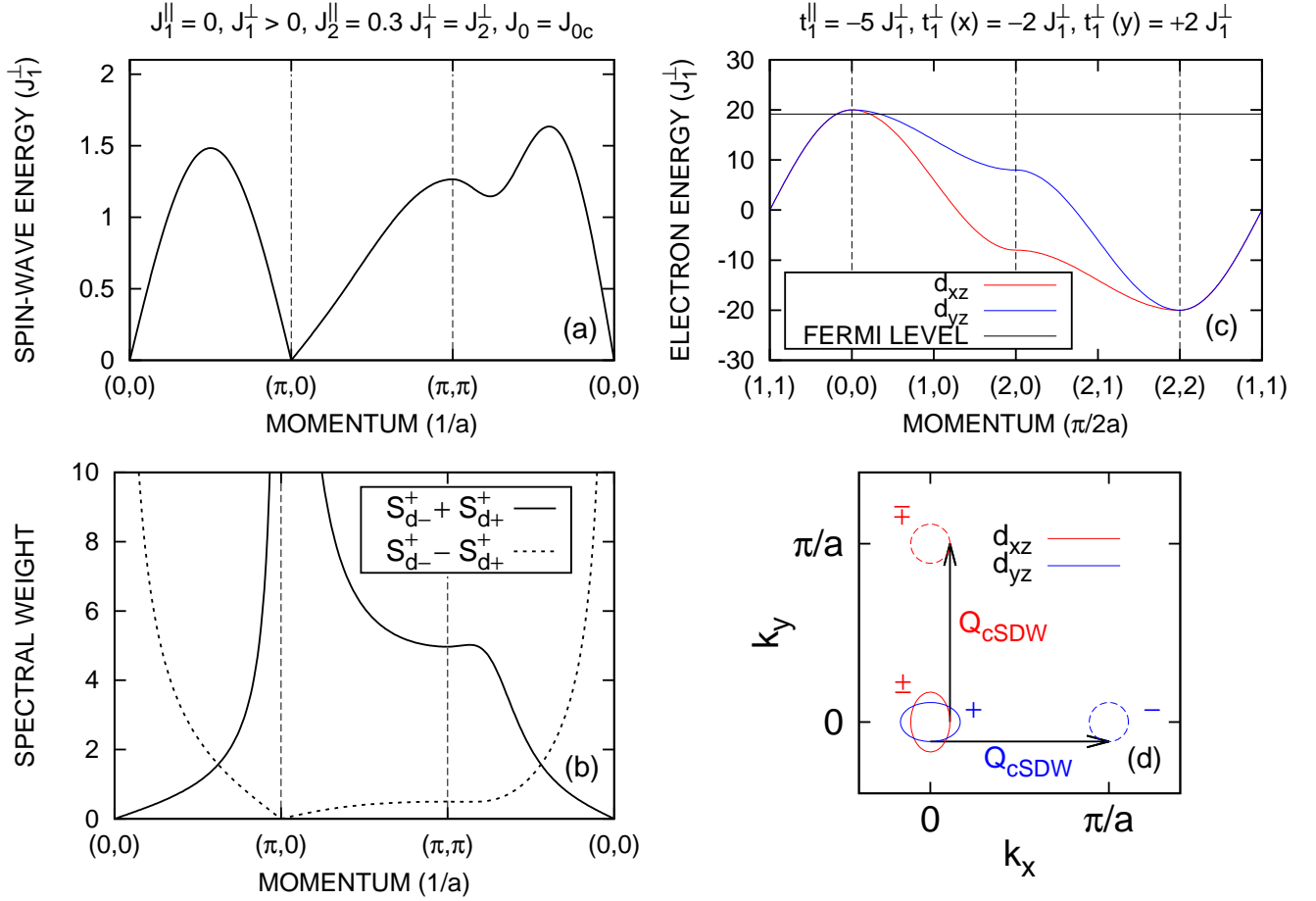}
\caption{The critical low-energy spectrum (a)-(b) of the
two-orbital Heisenberg model that corresponds to Eq. (\ref{tJ})  
in the linear spin-wave approximation (ref. \cite{jpr_10}).
The Hund coupling is set to $-J_{0c} = 0.8\, J_1^{\perp}$.
True ($q_0 = 0$) and hidden ($q_0 = \pi$) spinwaves are degenerate.
Shown also is 
(c) the band structure
and
(d) emergent nesting (ref. \cite{jpr_mana_pds_14})
for the two-orbital $t$-$J$ model, Eq. (\ref{tJ}).
The Fermi-level in panel (c) corresponds to 
a non-optimal half metal phase
obtained from Schwinger-boson-slave-fermion meanfield theory
at subcritical Hund coupling (ref. \cite{jpr_mana_pds_14}).}
\label{FS}
\end{figure}

After tuning the Hund coupling to the QCP,
and in the absence of hybridization between the $d_{xz}$ and $d_{yz}$ orbitals,
we find that the groundstate is a spin-$0$ Cooper pair
with primarily $S^{+-}$ symmetry.
It is in a bonding superposition of
orbitally ordered $d_{yz}^{2(+-)}$ and $d_{xz}^{2(+-)}$ singlet pairs
that alternate in sign between respective hole and electron pockets. 
(Cf. ref. \cite{sup_mat_3}.)
The $S^{+-}$ hole pair is
well separated from a continuum of states,
but close by in energy to an anti-bonding $D^{+-}$ Cooper pair.
(See Fig. \ref{FS}d, blue versus red.)
The macroscopic superconductor will exhibit an
internal Josephson effect between the two species of 
orbitally ordered Cooper pairs\cite{josephson_62,anderson_64,leggett_66,leggett_75}.
We thereby predict a neutral spin-$0$ collective mode in
the macroscopic superconducting state
that exhibits orbital pair oscillations.  
It couples directly to orthorhombic shear\cite{yoshizawa_12,yoshizawa_simayi_12}.

Last,
the exact energy difference separating the $D^{+-}$ and the $S^{+-}$
hole-pair states
tends to zero linearly with increasing Hund coupling
near the QCP. (See Fig. \ref{dE0vsJ0}b.)
By comparison with mean-field theory and
exact results for one mobile hole\cite{jpr_mana_pds_14},
we argue in the Discussion section that the quantum critical state\cite{qcp},
where $S^{+-}$ and $D^{+-}$ Cooper pairs become degenerate,
coincides with the Lifshitz transition, 
where electron bands at cSDW momenta
first rise above the Fermi level\cite{malaeb_prb_12,budko_prb_13,hodovanets_prb_14}.
This picture implies remnant Cooper pairs of opposite sign on the emergent
electron bands that lie above the Fermi level at subcritical Hund coupling. 
(See Figs. \ref{emerge}b and \ref{s+-d+-}.)
A recent analysis of a phenomenological model 
for over-doped Ba$_{1-x}$K$_x$Fe$_2$As$_2$ 
reaches a similar conclusion\cite{bang_14}.
The present theory also then predicts that the in-gap $s$-to-$d$-wave collective mode
observed recently by Raman spectroscopy in optimally doped Ba$_{1-x}$K$_x$Fe$_2$As$_2$
will also be observed at the Lifshitz transition at higher doping\cite{raman_13,raman_14}.

\section{Local-Moment Model}
The Hamiltonian for the two-orbital $t$-$J$ model on 
a square lattice of iron atoms reads\cite{jpr_mana_pds_11,jpr_mana_pds_14}
\begin{eqnarray}
H &=& \sum_{\langle i,j \rangle} \bigl[
     - (t_1^{\alpha,\beta} {\tilde c}_{i, \alpha,s}^{\dagger} {\tilde c}_{j,\beta,s} + {\rm h.c.})
     + J_1^{\alpha,\beta}\bigl( {\bm S}_{i, \alpha} \cdot {\bm S}_{j, \beta}
                               + {1\over 4}n_{i, \alpha} n_{j, \beta}\bigr)\bigr] + \nonumber \\
    && \sum_{\langle\langle i,j \rangle\rangle}
J_2^{\alpha,\beta}\bigl( {\bm S}_{i, \alpha} \cdot {\bm S}_{j, \beta}
                          + {1\over 4}n_{i,\alpha} n_{j,\beta}\bigr) +
   \sum_i \bigl( J_0 {\bm S}_{i,d-}\cdot {\bm S}_{i,d+}
    +  U_0^{\prime} {\bar n}_{i,d+} {\bar n}_{i,d-} \bigr). \nonumber \\
\label{tJ}
\end{eqnarray}
Above, ${\tilde c}_{i, \alpha,s} = b_{i, \alpha,s} f_{i,\alpha}^{\dagger}$ is the
destruction operator for an electron of spin $s$, at site $i$, in orbital $\alpha$.
Also,
${\bm S}_{i,\alpha} =
{1\over 2} \sum_{s,s{\prime}}
{\tilde c}_{i,\alpha,s}^{\dagger}
{\bm \sigma}_{s,s^{\prime}} {\tilde c}_{i,\alpha,s^\prime}$
is the spin operator in units of $\hbar$,
$n_{i,\alpha} = \sum_s {\tilde c}_{i, \alpha,s}^{\dagger} {\tilde c}_{i, \alpha,s}$
measures the net occupation per site-orbital,
while ${\bar n}_{i,\alpha} = 1 - n_{i,\alpha}$ counts holes instead.
Electrons live on the $d+ = d_{(x+iy)z}$ and/or the $d- = d_{(x-iy)z}$ orbitals.
Repeated orbital and spin indices
in the hopping and Heisenberg exchange terms above
are summed over.
These terms in the Hamiltonian (\ref{tJ})
are in turn summed over
nearest neighbor  and next-nearest neighbor links,
$\langle i,j \rangle$ and $\langle\langle i,j \rangle\rangle$.
Double occupancy at a site-orbital is projected out by enforcing the constraint
\begin{equation}
1 = b_{i,\alpha,\uparrow}^{\dagger} b_{i,\alpha,\uparrow}
+ b_{i,\alpha,\downarrow}^{\dagger} b_{i,\alpha,\downarrow}
+ f_{i,\alpha}^{\dagger} f_{i,\alpha},
\label{constraint}
\end{equation}
where $b_{i,\alpha,\uparrow}$ and $b_{i,\alpha,\downarrow}$ are the destruction operators for
a pair of Schwinger bosons,
and where $f_{i,\alpha}$ is the destruction operator for 
a spinless slave fermion\cite{kane_89,auerbach_larson_91}.
In order to  reduce finite-size effects,
we have added a {\it repulsive} interaction to the Heisenberg exchange terms above.
The net interaction between nearest neighbors ($n=1$)
and between next-nearest neighbors ($n=2$)
is thereby pure spin exchange: 
${1\over2} J_{n}^{\alpha,\beta} P_{i, \alpha; j, \beta}$.
Here, the operator $P_{i, \alpha; j, \beta}$ exchanges spins on site-orbitals
$(i, \alpha)$ and $(j, \beta)$.
Observe the invariance of the Hamiltonian under
the following internal global gauge transformation:
${\tilde c}_{i, d\pm, s} \rightarrow e^{\pm i\delta_0}{\tilde c}_{i, d\pm, s}$
and
$t_1^{d\pm d\mp} \rightarrow e^{\pm 2i\delta_0}t_1^{d\pm d\mp}$.
It is equivalent to a rotation of the orbital coordinates $(x,y)$ by an angle $\delta_0$.
The $t$-$J$ model (\ref{tJ}) is then invariant under 
an arbitrary rotation of the two orbitals about the $z$ axis.

Below, we will provide evidence that the two-orbital $t$-$J$ model
is unstable to the formation of local intra-orbital Cooper pairs
in the vicinity of a QCP that separates cSDW order from hidden magnetic order. 
The second column in Table \ref{table1} coincides with the parameter range of the model.
We will further show that 
an internal Josephson effect exists, specifically,
between $d_{yz}$-$d_{yz}$ and $d_{xz}$-$d_{xz}$
singlet Cooper pairs.
Before doing so, however, we first demonstrate how
emergent electron bands begin to nest with hole Fermi surface pockets at the QCP.

\section{Coherent Hole Bands and Emergent Electron Bands}
Semi-classical calculations of
the Heisenberg model that corresponds to (\ref{tJ})
at half filling
find a QCP
that separates a cSDW at strong Hund coupling
from a hidden antiferromagnet
at weak Hund coupling
if off-diagonal frustration is present\cite{jpr_10}: e.g.
$J_1^{\parallel} = 0$, $J_1^{\perp} > 0$, and  $J_2^{\parallel} = J_2^{\perp} > 0$.
The antiferromagnetic sublattices of 
the hidden-order state are the  $d+$ and $d-$ orbitals.
In particular,
the groundstate at large electron spin $s_0$
is the spin texture $\nwarrow_{d-}\searrow_{d+}$.
Ideal hopping of holes within an antiferromagnetic sublattice,
$t_1^{\parallel} < 0$ and $t_1^{\perp} = 0$,
leaves such hidden magnetic order intact.
In such case,
a mean-field treatment of (\ref{tJ}) and (\ref{constraint})
at Hund coupling below the QCP
reveals a half metal
with doubly-degenerate Fermi surface pockets
at zero 2D momentum
that are circular and hole-type\cite{jpr_mana_pds_11,jpr_mana_pds_14}. 
The second column in Table \ref{table1} summarizes the above.
Below, we demonstrate that
emergent electron bands at cSDW momenta
are also predicted within the mean-field approximation.

\begin{table}
\begin{center}
\begin{tabular}{|c|c|c|}
\hline
filling, bands &
$J_1^{\parallel} < J_1^{\perp}$ & $J_1^{\parallel} > J_1^{\perp}$ \\
\hline
half filling, none   &
hidden ferromagnet: $(\pi,0,0)$ & hidden N\'eel: $(\pi,\pi/a,\pi/a)$ \\
hole dope, hole bands at $\Gamma$ &
hidden half metal, FS pockets at $\Gamma$ & nested cSDW metal? \\
\hline
\end{tabular}
\caption{Groundstate of two-orbital $t$-$J$ model (\ref{tJ}).
Hund coupling is tuned to the QCP at half filling, which separates a cSDW at strong Hund
coupling from hidden magnetic order at weak Hund coupling (ref. \cite{jpr_10}).
The $3$-vector $(\pi, k_x, k_y)$ describes the hidden magnetic order.}
\label{table1}
\end{center}
\end{table}
\begin{figure}
\includegraphics[scale=0.65, angle=0]{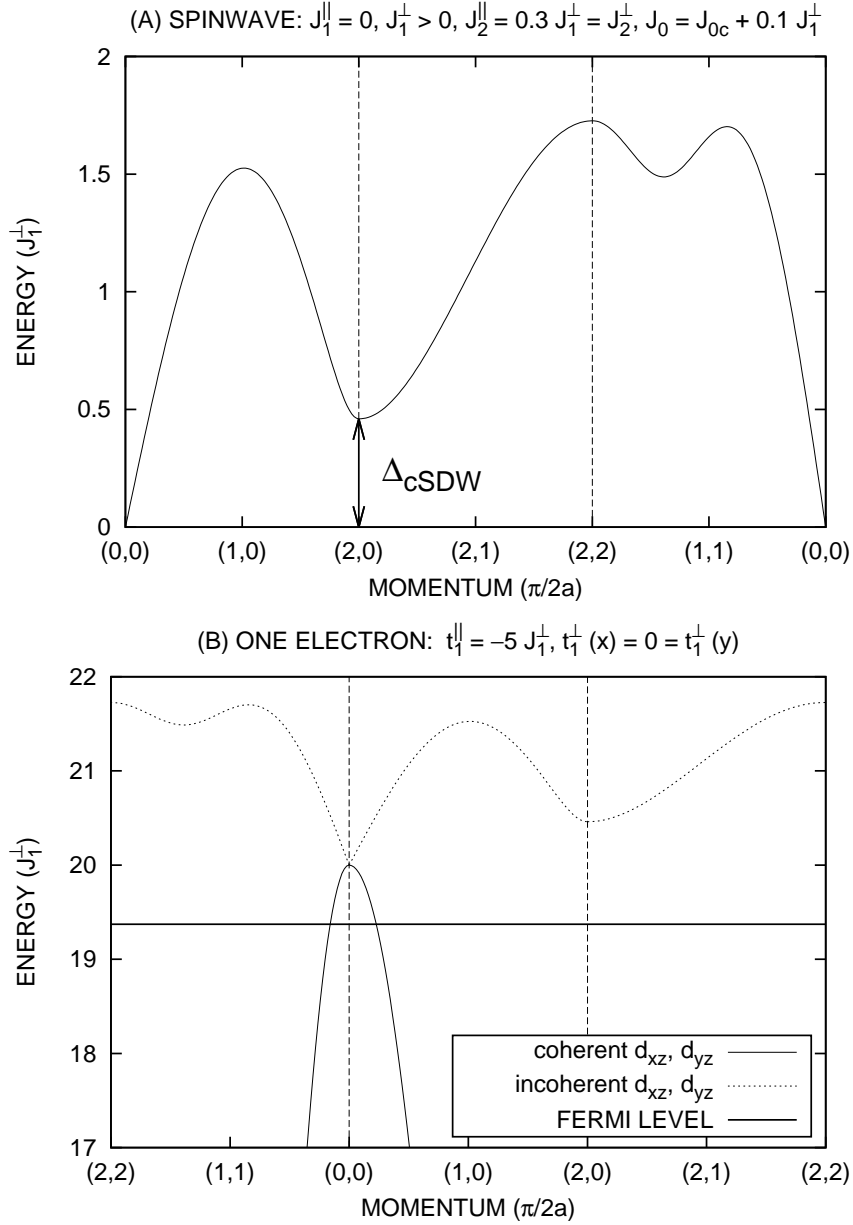}
\caption{(a) The energy-dispersion relation for spinwaves (Schwinger bosons)
in the half-metal state near the QCP and
(b) the corresponding imaginary part of the one-electron propagator near half filling,
Eq. \ref{emergent_electron},
at site-orbital concentration $x=0.01$.
Intrinsic broadening due to incoherent contributions in Eq. \ref{G}
is not shown. (See text.)}
\label{emerge}
\end{figure}
%

Electronic structure is revealed by
the propagator for one electron 
at energy $\omega$,
at 3-momentum $k = (k_0,{\bm k})$.
Here the quantum numbers $k_0 = 0$ and $\pi$ represent
bonding and anti-bonding superpositions of the $d-$ and $d+$ orbitals,
which are respectively the $d_{xz}$ and the $(-i)d_{yz}$ orbitals.
The latter are good quantum numbers in the absence of hybridization,
which is the case throughout.
Within a mean-field approximation for the half-metal state,
where the constraint against double occupancy (\ref{constraint}) 
is enforced on average over the bulk,
the one-electron propagator
is given by the expression\cite{jpr_mana_pds_11}
\begin{eqnarray}
G(k,\omega) = {1\over{\cal N}}\sum_q &\Bigl[&({\rm cosh}\, \theta_{q+k})^2
{{n_B[\omega_b (q+k)] + n_F[\varepsilon_f (q)]}\over{\omega - \omega_b(q+k) + \varepsilon_f(q)}}  \nonumber  \\
&&+({\rm sinh}\, \theta_{q+k})^2
{{n_B[\omega_b (q+k)] + n_F[-\varepsilon_f(q)]}\over{\omega + \omega_b(q+k) + \varepsilon_f(q)}}\Bigr] .
\label{G}
\end{eqnarray}
Above,
$\omega_b  = (\Omega_{\parallel}^2 - \Omega_{\perp}^2)^{1/2}$
is the energy dispersion of the spin-carrying Schwinger bosons,
while $\varepsilon_f({\bm k}) = 8 s_0 t_1^{\parallel}  \gamma_1({\bm k}) - \mu$
is the energy dispersion of the charge-carrying slave fermions
measured with respect to the chemical potential $\mu$.
Both are degenerate in the orbital channels $k_0 = 0$ and $\pi$.
Here\cite{jpr_mana_pds_11},
\begin{eqnarray*}
\Omega_{\parallel}(k) &=&
s_0\sum_{n=0,1,2} z_n J_n^{\prime\perp}
-4\sum_{n=1,2} (s_0 J_n^{\prime\parallel}+t_n^{\parallel} x)[1-\gamma_n({\bm k})] ,\\
\Omega_{\perp}(k) &=&
s_0 e^{i k_0} \sum_{n=0,1,2} z_n J_n^{\prime\perp} \gamma_n({\bm k}) ,
\end{eqnarray*}
and $\gamma_1 ({\bm k}) = {1\over 2} ({\rm cos}\, k_x a\, +  \,{\rm cos}\, k_y a)$.
Above, $\gamma_0 ({\bm k}) = 1$ and
$\gamma_2 ({\bm k}) = {1\over 2} ({\rm cos}\, k_+ a\,+\,{\rm cos}\, k_- a)$,
with $k_{\pm} = k_x \pm k_y$,
while $z_0 = 1$ and $z_1 = 4 = z_2$ are coordination numbers,
and while $J^{\prime} = (1-x)^2 J$.
Also above, $n_B$ and $n_F$ denote the Bose-Einstein and the Fermi-Dirac distributions,
while ${\cal N} = 2 N_{\rm Fe}$ denotes the number of site-orbitals
on the square lattice of $N_{\rm Fe}$ iron atoms.
Last, the coherence factors for Schwinger bosons that appear in (\ref{G})
are set by
${\rm cosh} \, 2\theta = \Omega_{\parallel} / \omega_b$ and
${\rm sinh} \, 2\theta = \Omega_{\perp} / \omega_b$.
Ideal
Bose-Einstein condensation (BEC)
of the Schwinger bosons in the Goldstone mode shown in Fig. \ref{emerge}a 
results in the following
coherent contribution to the electronic spectral function\cite{jpr_mana_pds_11}
at zero temperature and at large electron spin $s_0$:
${\rm Im}\, G_{\rm coh}(k,\omega) = s_0
{\pi} \delta[\omega + \varepsilon_f(k)]$.
It reveals degenerate hole bands
for $d_{xz}$ and $d_{yz}$ orbitals
centered at zero 2D momentum,
with circular Fermi surfaces of radius  $k_F a = (4\pi x)^{1/2}$.
Here, $x$ denotes the concentration of holes per iron site, per orbital.
The hole Fermi surface pockets at $\omega = 0$
are depicted by Fig. \ref{FS}d. 
Mass anisotropy has been added
to distinguish between the two orbitals.
At energies
above the Fermi level, $\omega > 0$,
the remaining contribution from incoherent excitations
${\rm Im}\, G_{\rm inc}(k,\omega)$
are exclusively due to
the first fermion term in (\ref{G}).
It is easily evaluated
in the limit near half-filling,
$k_F a\rightarrow 0$, at large $t/J$:
%
\begin{equation}
{\rm Im}\, G_{\rm inc}(k,\omega) \cong 
\pi x 
\Bigl({1\over 2} + {1\over 2}{\Omega_{\parallel}\over{\omega_b}}\Bigl|_{\bm k}\Bigr)
\delta[\omega - \epsilon_F - \omega_b({\bm k})].
\label{emergent_electron}
\end{equation}
Figure \ref{emerge}b displays the emergent electron bands
predicted above.
They lie $\epsilon_F + \Delta_{cSDW}$ above the Fermi level,
with degenerate minima at both cSDW wavenumbers
$(\pi/a){\bm{\hat x}}$ and $(\pi/a){\bm{\hat y}}$,
for both the $d_{xz}$ and $d_{yz}$ orbitals.
Here, $\epsilon_F = (2 s_0) |t_1^{\parallel}| (k_F a)^2$
is the Fermi energy,
and $\Delta_{cSDW}$ is the spin gap at cSDW wavenumbers 
shown by Fig. \ref{emerge}a.
It vanishes at the QCP as
$\Delta_{cSDW} \propto {\rm Re}\, (J_0 - J_{0c})^{1/2}$,
with a critical Hund coupling\cite{jpr_mana_pds_11}
\begin{equation}
- J_{0c} = 2 (J_1^{\perp} - J_1^{\parallel}) - 4 J_2^{\parallel} -
  (1-x)^{-2} s_0^{-1} 2  t_1^{\parallel} x .
\label{J_0c}
\end{equation}
The emergent bands exhibit intrinsic broadening in frequency,
which can be calculated outside the critical region, at large $t/J$:
$\Delta\omega\sim k_F|{\bm\nabla} \omega_b|_{\bm k}$.
It remains small at the previous mimima.
Notice
the divergence of the spectral weight in (\ref{emergent_electron})
at cSDW wavenumbers,
as the cSDW spin gap vanishes at the QCP.
It yields  the bound
$\Delta_{cSDW} > k_F v_0$
that  guarantees the
validity of (\ref{emergent_electron}).
Here, $v_0$ denotes the velocity of cSDW spinwaves at the QCP.
(See ref. \cite{jpr_mana_pds_11} and Appendix \ref{ppndx_spin_gap}.)
The same criterion results in the bound $|{\bm k}| > k_F$,
which guarantees the validity of (\ref{emergent_electron})
near ${\bm k} = 0$.
Last, Fig. \ref{emerge} displays how
the emergent electron bands predicted by (\ref{emergent_electron})
inherit the mass anisotropy of the spinwaves at cSDW momenta.

At one electron less than half filling,
exact calculations of
the two-orbital $t$-$J$ model (\ref{tJ})
over a $4\times 4$ lattice of iron atoms
obtain a low-energy spectrum
that compares well with the emergent electron bands (\ref{emergent_electron})
shown by Fig. \ref{emerge}b,
but in the absence of Hund's rule. 
(See ref. \cite{jpr_mana_pds_11}, Fig. 3a.)
Adding inter-orbital electron hopping, 
$t_1^{\perp}({\hat{\bm x}}) < 0$, with 
$t_1^{\perp}({\hat{\bm y}}) = -t_1^{\perp}({\hat{\bm x}})$,
breaks orbital degeneracy both in 
the coherent hole bands and in the emergent electron bands.
Exact results over a $4\times 4$ lattice of iron atoms
find that such inter-orbital hopping results 
in an energy splitting between the $d_{xz}$ and $d_{yz}$ hole states
at momenta $\pm (\pi/2a){\hat{\bm x}}$ and $\pm (\pi/2a){\hat{\bm y}}$
that reflects the  mass anisotropies depicted by Fig. \ref{FS}d.
(See ref. \cite{jpr_mana_pds_14}, Figs. 4 and 5.)
The exact results also find
emergent electron bands 
at cSDW wave numbers
${\bm Q}_{0} = (\pi/a){\bm{\hat y}}$ and ${\bm Q}_{\pi} =(\pi/a){\bm{\hat x}}$
with respective $d_{xz}$ and $d_{yz}$ orbital character\cite{jpr_mana_pds_14}.
They become degenerate with the doubly-degenerate groundstates at zero 2D momentum
at the QCP,
where Hund coupling is tuned to the critical
value $-J_{0c} = 1.733\, J_1^{\perp}$.
Heisenberg-exchange and electron-hopping parameters
coincide with those
set in Fig. \ref{FS}.
Last, the first-excited states for the specific orbital $k_0$
at neighboring momenta
${\bm Q}_{k_0} \pm (\pi/2a){\hat{\bm x}}$ and ${\bm Q}_{k_0} \pm (\pi/2a){\hat{\bm y}}$
are nearly degenerate in such case.
(See ref. \cite{jpr_mana_pds_14}, Fig. 5.)
This is consistent with emergent electron bands at cSDW momenta
with nearly isotropic effective masses near the QCP.

\section{Low-Energy Spectrum of Hole Pair}
The exact low-energy spectrum of a pair of holes that roam
over a periodic $4\times 4$ square lattice of iron atoms 
governed by the two-orbital $t$-$J$ model (\ref{tJ})
can be obtained numerically.
The total spin along the $z$ axis is constrained to $\sum S_z = 0$.
Quantum states are defined by
a given spin background over the entire lattice
combined with a pair of spin-up and spin-down site-orbitals designated as holes.
The former defines the ensemble of Schwinger bosons, which we treat in occupation space,
while the latter defines the pair of slave fermions, which we treat in first quantization.
Translation symmetry and 
a combination of spin-flip symmetry with slave-fermion exchange is also included,
as well as reflection symmetries that leave momenta invariant.
Due to the absence of hybridization among the $d_{xz}$ and $d_{yz}$ holes,
orbital swap symmetry $P_{d,{\bar d}}$ is further added to the list:
$d+\leftrightarrow d-$.
For example, 
including an even parity reflection about the $x$ axis,
an even parity spin-flip, plus even parity orbital swap
reduces the dimension of the Hilbert space 
with net momentum $\hbar (\pi/2a){\bm{\hat x}}$ to 601 878 172 states.
The ARPACK subroutine library is exploited
to obtain low-energy eigenstates via the Lanczos technique\cite{arpack}.
Also,
matrix-vector products are accelerated throughout by running parallel OpenMP threads.

\begin{figure}
\includegraphics[scale=0.65, angle=-90]{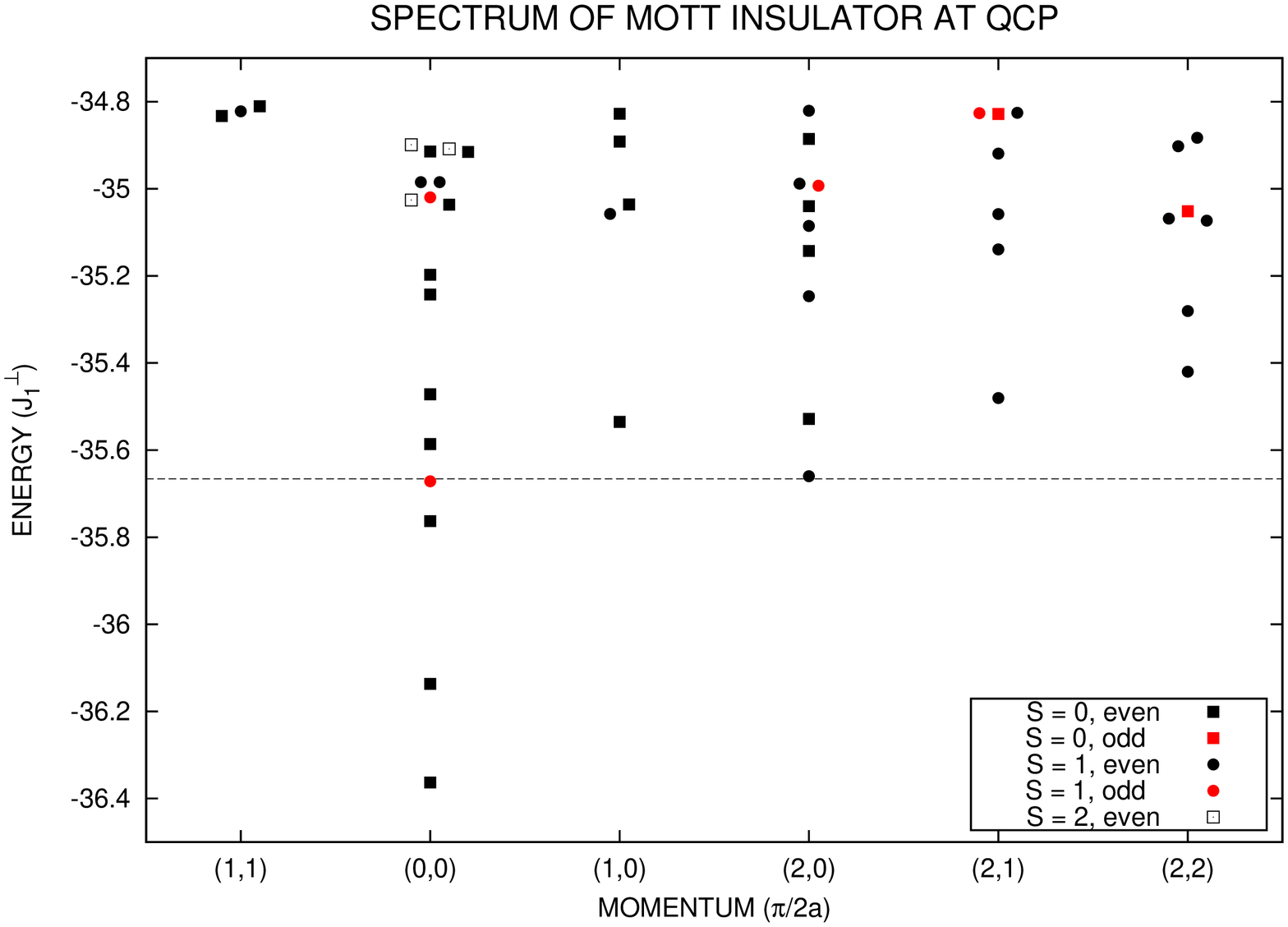}
\caption{The exact critical low-energy spectrum  of the
two-orbital Heisenberg model that corresponds to
the two-orbital $t$-$J$ model Eq. (\ref{tJ}) 
over a $4\times 4$ lattice (ref. \cite{jpr_10}).
Heisenberg exchange coupling constants coincide with those in Fig. \ref{FS}a,
while the Hund coupling is $-J_{0c} = 1.35\, J_1^{\perp}$.
Black and red energy levels are respectively even and odd under $P_{d,{\bar d}}$.
The horizontal dashed line marks the putative QCP: $\Delta_{cSDW} = 0$.}
\label{sw_spctrm_qcp}
\end{figure}

{\it Half Filling.}
In the absence of mobile holes,
the two-orbital $t$-$J$ model (\ref{tJ})
describes a frustrated antiferromagnetic insulator.
Figure \ref{sw_spctrm_qcp} shows 
the exact critical spectrum of 
the corresponding two-orbital Heisenberg model
over a $4\times 4$ lattice of iron atoms,
with exchange coupling constants
$J_1^{\parallel} = 0$,
$J_1^{\perp} > 0$, and
$J_2^{\parallel} = 0.3\, J_1^{\perp} = J_2^{\perp}$.
The Hund coupling is tuned to a critical value of
$-J_{0c} =  1.35\, J_1^{\perp}$,
at which point
true spinwaves at cSDW momenta that
have even parity under swap of the $d+$ and $d-$ orbitals
become degenerate with
hidden-order spinwaves at zero 2D momentum that have odd parity under such orbital swap.
The hidden-order spinwave (hSW) signals
long-range antiferromagnetic correlations across the
$d+$ and $d-$ orbitals
at wavenumber ${\bm Q} = 0$ \cite{jpr_10},
$\nwarrow_{d-}\searrow_{d+}$,
with order parameter
\begin{equation}
\langle S_{d-}^{+} ({\bm Q}) - S_{d+}^{+} ({\bm Q})\rangle
= i\hbar \sum_{\bm k} \langle 
{\tilde c}_{d_{xz},\uparrow}^{\dagger} ({\bm k}) {\tilde c}_{d_{yz},\downarrow}({\bm k}+{\bm Q})
-{\tilde c}_{d_{yz},\uparrow}^{\dagger} ({\bm k}) {\tilde c}_{d_{xz},\downarrow}({\bm k}+{\bm Q})\rangle .
\label{hidden_op}
\end{equation}
Here, ${\tilde c}_{o,s} ({\bm k})$
destroys a strongly correlated electron of spin $s$ in orbital $o$
that carries momentum $\hbar {\bm k}$.
The hidden magnetic order parameter (\ref{hidden_op}) is an orbital singlet,
which is odd (``red'') under $P_{d,{\bar d}}$,
and it manifestly probes {\it inter-orbital nesting}\cite{jpr_mana_pds_14}.
Such hidden magnetic order
becomes possible at weak intra-orbital Heisenberg exchange,
$J_1^{\parallel} < J_1^{\perp}$,
at subcritical Hund coupling.
Notice that the dispersion of low-energy spin-$1$ excitations
in the exact critical spectrum, Fig. \ref{sw_spctrm_qcp},
is qualitatively similar to the critical spectrum predicted by
the linear spin-wave approximation\cite{jpr_10},
Fig. \ref{FS}a.
The latter occurs at 
a critical Hund coupling 
$-J_{0c} = 2 (J_1^{\perp} - J_1^{\parallel}) - 4 J_2^{\parallel} = 0.8\, J_1^{\perp}$
that is $40\,\%$
smaller than the former exact result.
Last, the linear spin-wave approximation
and exact results over a $4\times 4$ lattice of iron atoms
indicate that the above QCP marks a second-order  quantum phase transition\cite{qcp}
between a cSDW at strong Hund coupling
and a hidden-order magnet (\ref{hidden_op}) at weak Hund coupling.
(See Appendix \ref{ppndx_spin_gap} and ref. \cite{jpr_10}.)

\begin{figure}
\includegraphics[scale=0.65, angle=-90]{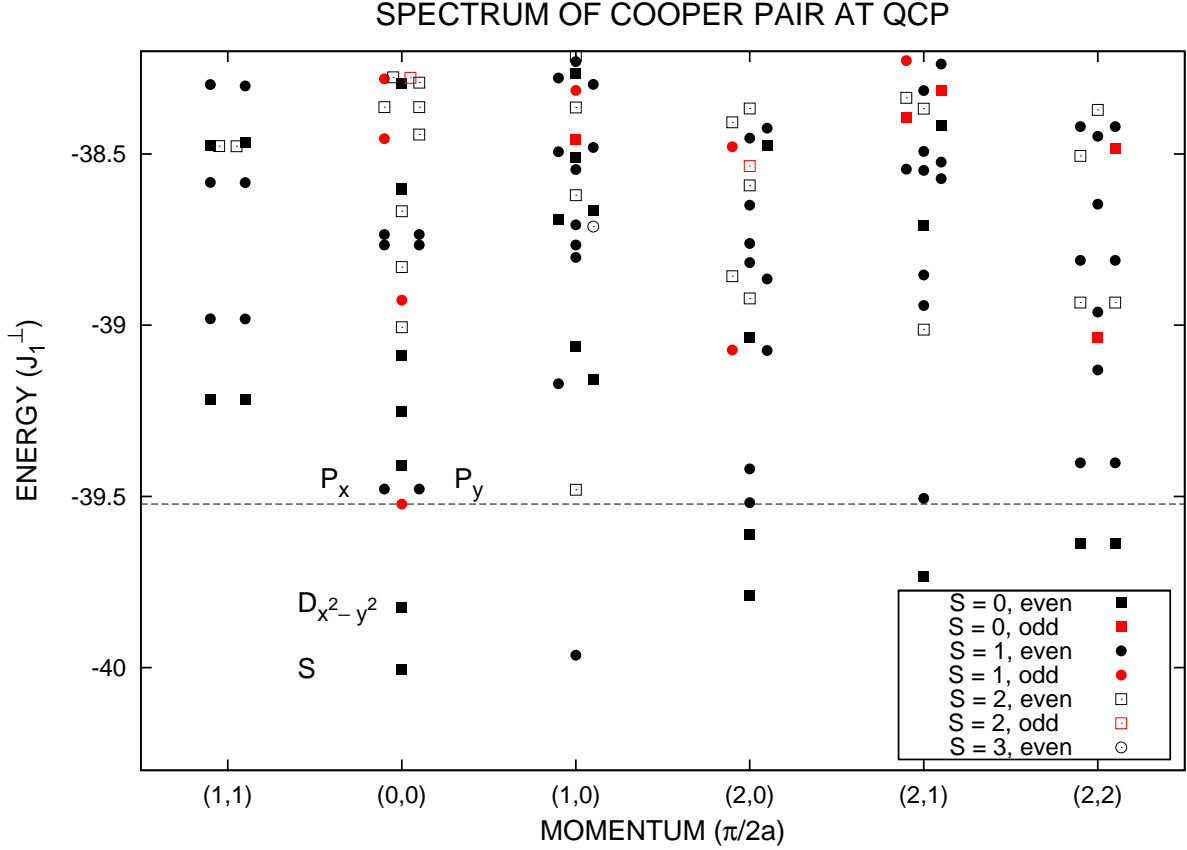}
\caption{The critical low-energy spectrum of the two-orbital $t$-$J$ model, 
Eq. (\ref{tJ})
plus the constant ${1\over 4} J_0 (N_{\rm Fe}-2)$,
for two holes roaming over a $4\times 4$  lattice. 
Heisenberg exchange and hopping parameters coincide with those in Fig. \ref{FS}.
Hund coupling is at the critical value
$-J_{0} = 2.25\, J_1^{\perp}$,
while inter-orbital on-site repulsion is set to
$U_0^{\prime} = {1\over 4}J_0 + 1000\, J_1^{\perp}$.
Black and red states are respectively even and odd under $P_{d,{\bar d}}$ .
Some points on the spectrum 
are artificially moved slightly off their quantized values along the
momentum axis for the sake of clarity.
The horizontal dashed line marks the putative QCP: $\Delta_{cSDW} = 0$.}
\label{spctrm}
\end{figure}

{\it Two Holes.}
Figure \ref{spctrm} displays the low-energy spectrum of two holes 
roaming over a $4\times 4$ lattice of spin-$1$ iron atoms
under periodic boundary conditions.  
The $t$-$J$ model
parameters are set to produce
hole bands at zero 2D momentum for 
non-interacting electrons (Fig. \ref{FS}c)
and cSDW spin order via magnetic frustration
when Hund's Rule is obeyed\cite{jpr_mana_pds_14}:
$t_1^{\parallel} = - 5 J_1^{\perp}$,
$t_1^{\perp} ({\bm{\hat x}}) = - 2 J_1^{\perp}$, 
$t_1^{\perp} ({\bm{\hat y}}) = + 2 J_1^{\perp}$,
along with the previous Heisenberg exchange coupling constants.
Further, 
the absence of next-nearest neighbor hopping imposes
conventional particle-hole symmetry in the hole spectrum.
It also turns off  hybridization between holes
in the $d_{xz}$ and the $d_{yz}$ orbitals.
Next,
the on-site hole-hole repulsion between the $d+$ and $d-$ orbitals
is set to a large value
$U_0^{\prime} = {1\over 4}J_0 + 1000\, J_1^{\perp}$.
Last, the ferromagnetic Hund's Rule exchange coupling constant
is  tuned to the critical value $J_{0} = - 2.25\, J_1^{\perp}$,
where true spin resonances at cSDW momenta
that have even parity under orbital swap
are degenerate
with a hidden-order spin resonance at zero 2D momentum
that has odd parity under orbital swap.
This defines a putative quantum critical point
that is realized at half filling in the limit of
large electron spin\cite{jpr_10} $s_0$.
(Cf. Figs. \ref{FS}a,b and  Fig. \ref{sw_spctrm_qcp}.)
The critical Hund coupling at half filling and at large $s_0$ is considerably smaller:
$-J_{0c} = 0.8\, J_1^{\perp}$.
The  larger Hund coupling in the present case of two mobile
holes is a result of 
the dominant intra-orbital hopping ($t_1^{\parallel}$) conspiring with
hidden anti-ferromagnetic order ($\nwarrow_{d-}\searrow_{d+}$)
to form a hidden half metal state
at weak Hund coupling\cite{jpr_10,jpr_mana_pds_11,jpr_mana_pds_14}.

\begin{figure}
\includegraphics[scale=0.65, angle=-90]{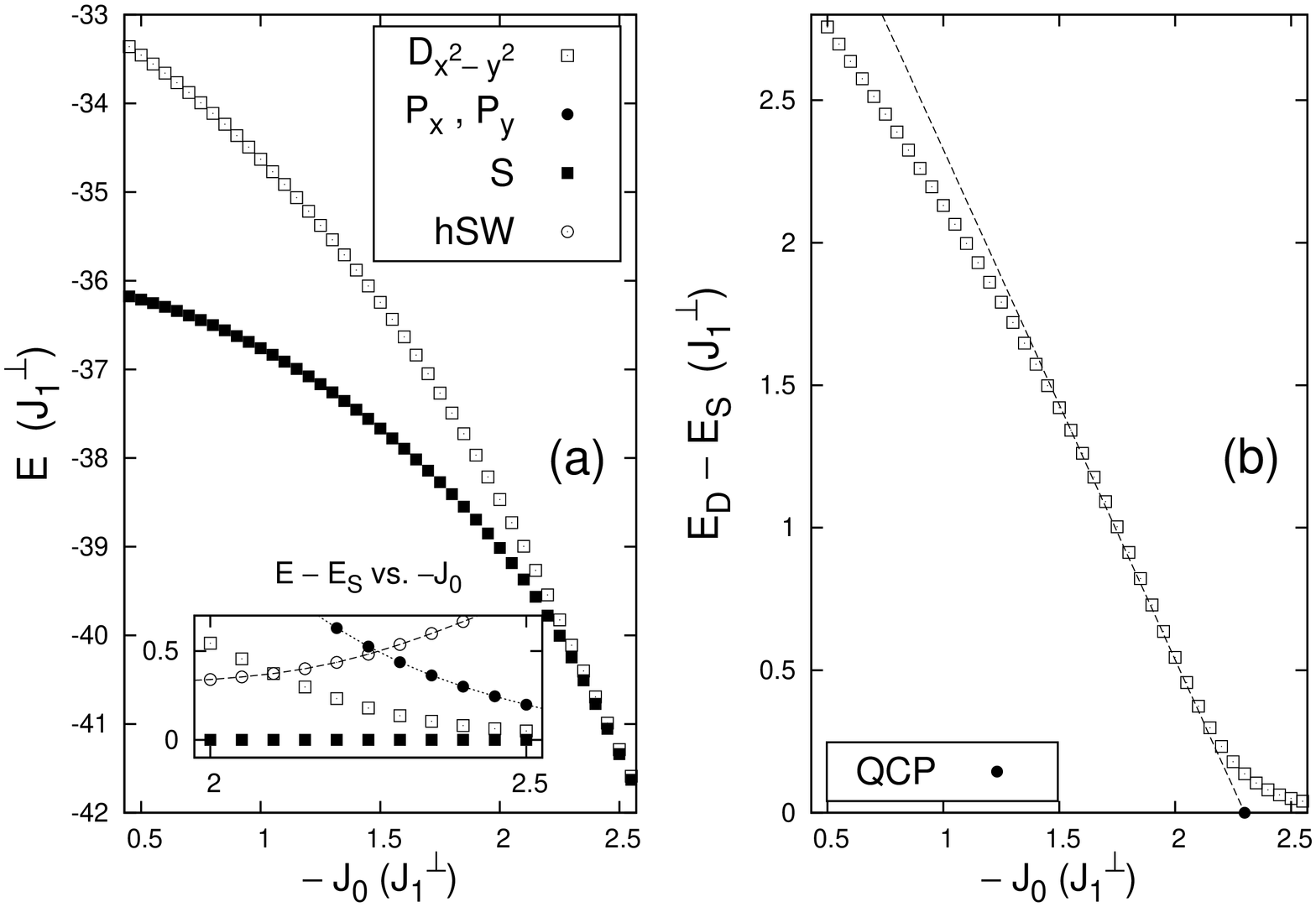}
\caption{(a) Exact energies for the groundstate $s$-wave pair state
and for the first-excited $d$-wave pair state versus Hund coupling.  
(b) Energy splitting versus Hund coupling.  The dashed line is a fit.
Model parameters $t$ and $J$ coincide with those in Fig. \ref{FS}.}
\label{dE0vsJ0}
\end{figure}

It is important to notice the groundstate and the zero-momentum excited state 
that lie below the horizontal dashed line in Fig. \ref{spctrm}.
Their reflection parities are listed in Table \ref{table2}.
Here, $R_{xz}$, $R_{yz}$ and $R_{x^{\prime}z}$ denote reflections about the $xz$ plane,
the $yz$ plane, and about the $(x+y)z$ plane.
The reflection parities indicate that the ground state is $s$-wave
and that the excited state is $d$-wave.
Figure \ref{dE0vsJ0} shows the evolution of 
the $s$-wave groundstate energy and of the $d$-wave excited-state energy
with increasing Hund coupling.
Notice how they merge at the putative QCP, yet avoid crossing.
The difference in energy versus Hund coupling
displays a prolonged inflection point there:
$E_{D} - E_{S} \propto J_0-J_{0c}$, with $-J_{0c}\cong 2.30 \, J_1^{\perp}$.
It suggests a QCP for a single pair of holes in 
the two-orbital $t$-$J$ model (\ref{tJ}) at the thermodynamic limit
that is consistent with the 
one predicted at half-filling by linear spin-wave theory
about hidden magnetic order\cite{jpr_10}.
Indeed, the QCP extracted from Fig. \ref{dE0vsJ0}b is close to
that specified by the degeneracy of the hidden spin-wave and
the true spin-wave excitations.
The latter is depicted by the horizontal dashed line in Fig. \ref{spctrm}.

\begin{table}
\begin{center}
\begin{tabular}{|c|c|c|c|c|c|c|}
\hline
no. & Hole-Pair State & $R_{xz}$ & $R_{yz}$ & $R_{x^{\prime}z}$ & $P_{d,{\bar d}}$ & spin \\
\hline
$0$  & $S$ & $+$ & $+$ & $+$ & $+$ & $0$  \\
$1$ & $D_{x^2-y^2}$ & $+$ & $+$ & $-$ & $+$ & $0$ \\
$2$ & hidden  spinwave & $-$ & $-$ & $-$ & $-$ & $1$ \\
$3a$ & $P_x$ & $+$ & $-$ & none & $+$ & $1$ \\
$3b$ & $P_y$ & $-$ & $+$ & none & $+$ & $1$ \\
\hline
\end{tabular}
\caption{Reflection parities, orbital-swap parity,
and spin of low-energy hole-pair states with zero net momentum
in order of increasing energy.
(See Fig. \ref{spctrm}.)}
\label{table2}
\end{center}
\end{table}

The inset to Fig. \ref{dE0vsJ0}a shows the dependence of low-energy levels 
at zero net momentum on Hund coupling in the vicinity of the QCP.
By contrast with the $s/d$-wave energy splitting,
the edge of the particle-hole continuum
in the exact two-hole spectrum at zero net momentum
does {\it not} collapse to the $s$-wave groundstate, 
nor to the $d$-wave excited state, as Hund coupling increases past the QCP.
Instead, the QCP coincides with the gap maximum marked
by the level crossing (dashed lines) in the inset to Fig. \ref{dE0vsJ0}a.
In other words, the quasi-particle gap ($E_b$)
remains nonzero at the QCP.

\begin{figure}
\includegraphics[scale=0.78, angle=0]{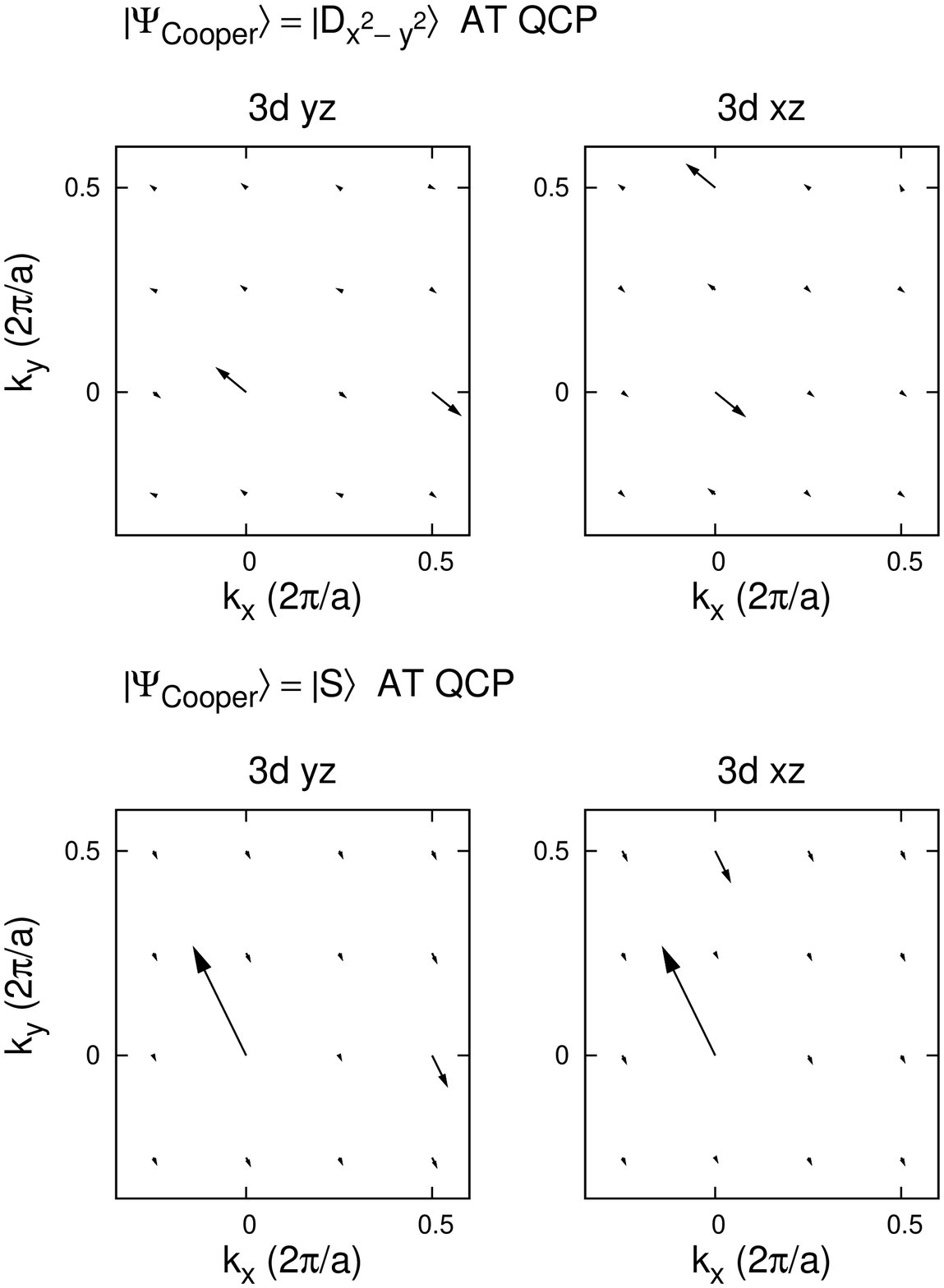}
\caption{The complex order parameter for superconductivity,
${\sqrt 2} i F(k)$,
of the $s$-wave groundstate 
and of the $d$-wave excited state in vector representation.
It is symmetrized with respect to both reflections about the principal axes.
Both the antiferromagnetic Mott insulator 
(Fig. \ref{sw_spctrm_qcp})
and the Cooper pair (Fig. \ref{spctrm}) are at the QCP.}
\label{s+-d+-}
\end{figure}

\section{Cooper Pairs and Collective Excitations}
The low-energy spectrum for the two-orbital $t$-$J$ model with two holes
displayed by Fig. \ref{spctrm} 
features nearly degenerate $s$-wave and $d$-wave hole-pair bound states.
Figure \ref{dE0vsJ0}b
strongly suggests that they become degenerate at a QCP in the thermodynamic limit.  
Below, we demonstrate that
these states are the principal members of a family of Cooper pairs (\ref{psi_phi})
characterized by a neutral spin-zero collective mode.

{\it Singlet Pairs.} The horizontal dashed line in Fig. \ref{spctrm} marks 
the degeneracy of the cSDW spin resonances with the hidden-order spin resonance
at zero 2D momentum.
At half filling,
the same degeneracy occurs 
at a QCP that separates cSDW order from hidden magnetic order
within the semi-classical approximation valid at large electron spin\cite{jpr_10}.
The dashed horizontal line in Fig. \ref{spctrm} lies at
the edge of a continuum of states with zero net momentum.
Two bound states exist below the continuum at zero net momentum:
an $s$-wave groundstate and a $d$-wave (second) excited state.
The former is even under a physical reflection about the $x$-$y$ diagonal
that includes a swap of the $d_{xz}$ and $d_{yz}$ orbitals,
while the latter is odd under it. (See Table \ref{table2}.)
Figure \ref{s+-d+-} depicts
the corresponding order parameters for superconductivity:
%
\begin{equation}
iF(k_0,{\bm k}) = \langle \Psi_{\rm Mott}|
{\tilde c}_{\uparrow}(k_0,{\bm k})^{\dagger}
{\tilde c}_{\downarrow}(k_0,-{\bm k})^{\dagger}|\Psi_{\rm Cooper}\rangle
\label{iF}
\end{equation}
times $\sqrt 2$,
with 
${\tilde c}_s(k_0,{\bm k}) = {\cal N}^{-1/2} \sum_i\sum_{\alpha=0,1}
e^{-i(k_0\alpha + {\bm k}\cdot{\bm r}_i)} {\tilde c}_{i,\alpha,s}$.
Here, ${\cal N} = 32$ is two times the number of iron atoms,
while the $d-$ and $d+$ orbitals $\alpha$ are enumerated by $0$ and $1$.
The bonding and anti-bonding superpositions of these orbitals,
$k_0 = 0$ and $\pi$, hence correspond to the $d_{xz}$ and $-i d_{yz}$ orbitals.
Also, $\langle \Psi_{\rm Mott}|$ denotes the critical antiferromagnetic state of the
corresponding Heisenberg model\cite{jpr_10}
at $-J_{0c} = 1.35\, J_1^{\perp}$. (See Fig. \ref{sw_spctrm_qcp}.)
The groundstate is primarily 
$| S^{+-} \rangle = {1\over{\sqrt 2}}|d_{yz}^{2(+-)}\rangle + {1\over{\sqrt 2}}|d_{xz}^{2(+-)}\rangle$,
with some admixture of 
$|S^{++}\rangle = {1\over{\sqrt 2}}|d_{yz}^{2(++)}\rangle + {1\over{\sqrt 2}}|d_{xz}^{2(++)}\rangle$:
$|S\rangle =
({\rm cos}\,\theta_0) |S^{+-}\rangle + ({\rm sin}\,\theta_0) |S^{++}\rangle$, 
where $\theta_0 = 27\,^{\circ}$.
Here, 
$|d_{yz}^{2(+-)}\rangle$ and $|d_{xz}^{2(+-)}\rangle$ are 
singlet Cooper pairs restricted to each orbital,
with equal and opposite values between
the corresponding hole and emergent electron pockets, 
while 
$|d_{yz}^{2(++)}\rangle$ and $|d_{xz}^{2(++)}\rangle$ are the same,
but with equal values at
the corresponding hole and emergent electron pockets. 
The former are sketched in Fig. \ref{FS}d (blue versus red),
and they are defined in Table \ref{table3}.
Figure \ref{s+-d+-}
also shows that the first-excited pair state is 
$| D^{+-} \rangle = {1\over{\sqrt 2}}|d_{yz}^{2(+-)}\rangle - {1\over{\sqrt 2}}|d_{xz}^{2(+-)}\rangle$.
The nature of the groundstate and first-excited pair states
is correctly captured by a three-state model analyzed in Appendix \ref{ppndx_3_pairs}.

And what is the diameter
of the $s$-wave and of the  $d$-wave Cooper pairs
identified in Fig. \ref{spctrm}? 
An orbital trace, $iF(0,{\bm k})-iF(\pi,{\bm k})$, 
of the principal $S^{+-}$ component 
in the $s$-wave groundstate shown by Fig. \ref{s+-d+-}
is approximated by 
${\rm cos}(k_x a)+{\rm cos}(k_y a)+{\rm cos}(k_+ a)+{\rm cos}(k_- a)$,
where $k_{\pm} = k_x \pm k_y$,
while an orbital trace of the $D^{+-}$ pair state
is approximated by
${\rm cos}(k_x a)-{\rm cos}(k_y a)$.
The $S^{+-}$ pair state is then approximately the superposition of 
all nearest neighbor and next-nearest neighbor pairs,
while the $D^{+-}$ pair state is approximately 
the difference of all $x$-axis-aligned
and of all $y$-axis-aligned nearest neighbor pairs.
Notice that these traces are consistent with
the component pair wavefunctions per orbital
listed in Table \ref{table3}.
The $s$-wave and $d$-wave bound states each thus fit in a unit cell.
This implies that the $4\times 4$ lattice studied here
is sufficiently big.
Macroscopic condensation into one of these pair states then
necessarily results in
a strong-coupling superconductor\cite{Seo_08,leggett_80,nozieres_schmitt-rink_85},
possibly of the short-range RVB type\cite{pBCS}.

\begin{table}
\begin{center}
\begin{tabular}{|c|c|c|}
\hline
Cooper Pair &
$d_{yz}$-$d_{yz}$ & $d_{xz}$-$d_{xz}$ \\
\hline
singlet $d_{yz}^{2(+-)}$  &
$(1+{\rm cos}\, k_y a) {\rm cos}\, k_x a $ & $0$ \\
singlet $d_{xz}^{2(+-)}$ &
$0$ & $(1+{\rm cos}\, k_x a) {\rm cos}\, k_y a $ \\
\hline
\end{tabular}
\caption{The pair amplitudes (\ref{iF}) of orbitally ordered states at the QCP.}
\label{table3}
\end{center}
\end{table}

{\it Orbital Pair  Oscillations.}
Following refs. \cite{josephson_62}, \cite{anderson_64}, and \cite{leggett_66},
an internal Josephson effect links the $s$-wave and $d$-wave pair states
identified by Fig. \ref{s+-d+-},
where holes condense into a dynamical Cooper pair state of the form
${1\over{\sqrt 2}} e^{i\phi(d_{yz})} |d_{yz}^{2(+-)}\rangle +
{1\over{\sqrt 2}} e^{i\phi(d_{xz})} |d_{xz}^{2(+-)}\rangle$.
Here, we have ignored the relatively small $S^{++}$ contribution of the $s$-wave
pair state evident in Fig. \ref{s+-d+-}.
A three-state model of the Cooper pairs in question finds that this is possible
sufficiently near the QCP. (See Appendix \ref{ppndx_3_pairs}.)
Notice that the probability of finding a pair of holes in either
a $|d_{yz}^{2(+-)}\rangle$ state or in a $|d_{xz}^{2(+-)}\rangle$ state
is equal to $1/2$ for the proposed pair wavefunction, as demanded by 2D isotropy.
An explicit Bardeen-Cooper-Schrieffer (BCS) wave function for
the macroscopic superconductor that projects out
double occupancy of electrons\cite{pBCS} per iron site-orbital
as well as double-occupancy of holes per iron atom
can be written down\cite{jpr_haas_15}.
It has the form\cite{anderson_64}
$|\Psi_{pBCS}\rangle = \sum_{N=0}^{\infty} c_N  e^{i N\phi}|\Psi_N\rangle$,
where
$|\Psi_N\rangle$ represents
$2N$ mobile holes condensed into the Cooper pair
\begin{equation}
|\psi_{J^{\prime}}(\phi_-)\rangle = 
{e^{+i\phi_- / 2}\over{\sqrt 2}} |d_{yz}^{2(+-)}\rangle +
{e^{-i\phi_- / 2}\over{\sqrt 2}} |d_{xz}^{2(+-)}\rangle ,
\label{psi_phi}
\end{equation}
%
and where $c_N$ are real constants that are sharply peaked at the mean number of pairs.
Above,
$\phi_- = \phi(d_{yz}) - \phi(d_{xz})$ and
$\phi = {1\over 2}[\phi(d_{yz}) + \phi(d_{xz})]$.
Orbital order 
$N_- = N(d_{yz}) - N(d_{xz})$
is sharply peaked at zero per
the binomial distribution at macroscopically large\cite{heuristic_pBCS}
$N = N(d_{yz}) + N(d_{xz})$.
Here, $N(o)$ counts the number of Cooper pairs in orbital $o$.
The phase of
the orbitally ordered pairs, $\phi(o)$,
is canonically conjugate to it.
We therefore have commutation relations
$[{1\over 2}\phi_-, N_-] = i = [\phi, N]$.
The remaining commutators vanish.  

An internal Josephson effect between the two species
of orbitally ordered Cooper pairs,
$|d_{yz}^{2(+-)}\rangle$ and $|d_{xz}^{2(+-)}\rangle$,
is then predicted by the following hydrodynamic Hamiltonian:
\begin{equation}
H_{\rm orb} =
V\Biggl[{1\over{2\chi_{\rm orb}}}\Biggl({N_-\over V}\Biggr)^2 -
{\mu_-\over 2} {N_-\over V} - 
e_{J^{\prime}}\, {\rm cos}(\phi_-)\Biggr] ,
\label{h_hydro}
\end{equation}
where $\mu_- = \mu(d_{yz}) - \mu(d_{xz})$
is the difference in chemical potential between the two orbitals.
Also, $V$ denotes the area of an iron-pnictide layer.
Above, $\chi_{\rm orb}$ is the susceptibility for orbital order
of the Cooper pairs at zero temperature.
It attains the normal-state value 
equal to one half the density of states
in the hydrodynamic regime\cite{leggett_takagi_77}.
Also, $e_{J^{\prime}}$
is one half the difference in condensation energy density
between the $S^{+-}$ state and the $D^{+-}$ state.
It is given by
$e_{J^{\prime}} \cong {1\over 2}(N/V)(E_D-E_S)$ in the regime of local Cooper pairs.
Notice from (\ref{psi_phi}) and (\ref{h_hydro})
that the equilibrium groundstate is $S^{+-}$,
and that it shows no orbital order:
$\phi_- = 0$ and $N_- = 0$ at $\mu_- = 0$.  
The $D^{+-}$ pair state,
on the contrary, is in unstable equilibrium:
$\phi_- = \pi$ and $N_- = 0$ at $\mu_- = 0$.  
Hydrodynamic equilibrium is therefore consistent
with the $s$-wave groundstate found by the previous exact
calculations at subcritical Hund coupling.
(See Fig. \ref{dE0vsJ0}.)
Orbital ordering of the macroscopic superconductor
is then governed by standard dynamical equations\cite{anderson_64}:
$\hbar {1\over 2}\dot\phi_- = \chi_{\rm orb}^{-1} (N_-/V) - {1\over 2}\mu_-$
and
$\hbar {1\over 2}\dot N_-/V = - e_{J^{\prime}}\, {\rm sin}(\phi_-)$,
while 
$N$, $\phi$ and $\mu = {1\over 2}[\mu(d_{yz}) + \mu(d_{xz})]$ remain constant.

The previous dynamical equations imply small oscillations in $N_-$ and $\phi_-$
about zero that are $90\,^{\circ}$ out of phase,
at a natural frequency $\omega_{J^{\prime}}$ that is related to 
the internal Josephson tunneling by\cite{leggett_66}
$(\hbar\omega_{J^{\prime}})^2 = 
4 e_{J^{\prime}} / \chi_{\rm orb}$.
The collective oscillation is undamped as long as it lies
inside the quasi-particle energy gap:
$\hbar\omega_{J^{\prime}} < E_b$.
Figure \ref{dE0vsJ0}b suggests
that $\omega_{J^{\prime}}$ collapses to zero at the QCP and/or at half-filling
as the product $(N/V)^{1/2} \cdot {\rm Re}\, (J_0 - J_{0c})^{1/2}$.
Unlike the $s/d$-wave energy splitting
in the exact spectrum, Fig. \ref{spctrm},
the energy gap that separates the
$S^{+-}$ groundstate from the edge of the particle-hole continuum 
does {\it not} tend to zero at the QCP.
(See the dashed lines in the inset to Fig. \ref{dE0vsJ0}a.)
The neutral spin-$0$ collective mode is hence observable
in the quantum critical region
or near half filling.
Last, orthorhombic shear strain
couples to this internal Josephson effect through
the electron-phonon interaction\cite{yoshizawa_12,yoshizawa_simayi_12}:
${1\over 2}\mu_- = \Xi_{u^{\prime}} {1\over 2} (\partial u_x / \partial x - \partial u_y / \partial y)$,
where $\Xi_{u^{\prime}}$ is a deformation potential, 
and where ${\bm u}$ is the displacement field of the iron atoms.
Level repulsion between the spectrum of orthorhombic phonons
and the above neutral spin-$0$ collective mode
is then possible if the Debye frequency
in the iron-pnictide superconductor
is larger than  $\omega_{J^{\prime}}$.
The Debye frequency is typically comparable to the energy gap 
in iron-pnictide superconductors\cite{lee_10}.
Hence, the collective mode couples strongly 
to such orthorhombic phonons
in the quantum critical region
or near half filling.

{\it Triplet Pairs.} The low-energy spectrum for two holes shown by Fig. \ref{spctrm}
also contains a degenerate pair of spin-$1$ states
with no net momentum
that lie at the edge of the continuum of states.
One is a $P_x$ state,
with even parity under a reflection about
the $x$ axis and odd parity under a reflection about the $y$ axis,
while the other is a $P_y$ state,
with these parities reversed.
(See Table \ref{table2}.)
This symmetry along with orbital order
is revealed by the
order parameter for superconductivity (\ref{iF})
of the $P_x$ state, for example:
${\sqrt 2}iF(k_0,{\bm k}) \cong 0.056\, {\rm sin} (k_x a)$ for $k_y$ near $0$ 
in the $d_{yz}$-$d_{yz}$ pair channel,
and ${\sqrt 2}iF(k_0,{\bm k}) \cong  0$ otherwise.
Here, ${\sqrt 2}iF(k)$ has been symmetrized with respect to a reflection about the $x$ axis.
Now notice the spin-$1$ state at momentum $\hbar (\pi/2a){\bm{\hat x}}$ in Fig. \ref{spctrm}
that is nearly degenerate with the groundstate.  
We propose that the former $P$ states are degenerate triplet Cooper pairs,
and that the latter spin-$1$ state is a remnant
of the corresponding Leggett mode\cite{leggett_75}.
Note that the possibility of an orbital singlet is excluded here.
At the QCP, it coincides instead with the hidden-order spin resonance 
at zero 2D momentum.
(See the ``red'' states in Fig. \ref{spctrm}.)

\section{Discussion}
Below, 
we describe the emergent nature of 
the $S^{+-}$ Cooper pair exhibited by Fig. \ref{s+-d+-}
and the implications of the QCP shown by  Fig. \ref{dE0vsJ0}b
on the collective mode.
We also argue that the $d$-wave pair state remains higher in energy than
the $s$-wave pair state at dilute concentrations of charge carriers,
and we survey the nature of the two-hole spectrum, Fig. \ref{spctrm},
at net momentum.

\subsection{Lifshitz Transition versus Quantum Critical Point}
Schwinger-boson-slave-fermion meanfield theory for the hidden half-metal state
predicts that the emergent electron bands shown by Fig. \ref{emerge}b
first cross the Fermi level at cSDW momenta when the energy difference
$\Delta_{cSDW}$ between cSDW spinwaves and 
the hidden-order spinwave at zero 2D momentum
vanishes\cite{jpr_mana_pds_11}. (See Fig. \ref{emerge}a.)  It therefore predicts that the
former Lifshitz transition coincides with the latter QCP
shown by Fig. \ref{FS}a at half filling.
Exact results on a $4\times 4$ lattice of iron atoms 
find that the putative QCP ($\Delta_{cSDW} = 0$) 
occurs at Hund coupling $-J_{0c} = 1.35\, J_1^{\perp}$ in 
the absence of mobile holes (Fig. \ref{sw_spctrm_qcp}),
while the Lifshitz transition occurs at $-J_{0c} = - 1.73\, J_1^{\perp}$
in the case of one mobile hole\cite{jpr_mana_pds_14}.  
In the case of two mobile holes, the putative QCP,
which is marked by the dashed horizontal line in Fig. \ref{spctrm},
lies at the critical Hund coupling $-J_{0c} = 2.25\, J_1^{\perp}$.
The monotonic increase of $-J_{0c}$ with the mobile hole concentration per orbital $x$
is consistent with the linear increase (\ref{J_0c}) of $-J_{0c}$ with $x$
obtained within the meanfield approximation\cite{jpr_mana_pds_11}.
Finally, 
the dependence
of the energy splitting between
the $S^{+-}$ hole-pair groundstate and the $D^{+-}$ hole-pair excited state 
on Hund coupling shown by Fig. \ref{dE0vsJ0}b
for a $4\times 4$ lattice of spin-$1$ iron atoms
suggests a QCP in the thermodynamic limit
at which these states become degenerate.
It occurs at $-J_{0c} = 2.30\, J_1^{\perp}$,
which is close to the putative QCP.
The $d_{yz}^{2(+-)}$ and $d_{xz}^{2(+-)}$ Cooper pairs
that are listed in Table \ref{table3} become degenerate boundstates
in such case, however.
This strongly suggests that the latter QCP
for a single pair of mobile holes
coincides with the Lifshitz transition exhibited by one mobile hole
and predicted by meanfield theory\cite{jpr_mana_pds_14}.
Comparison with Fig. \ref{s+-d+-} and the inset to Fig. \ref{dE0vsJ0}a
further also strongly suggests remnant Cooper pairs of opposite sign
on the emergent electron bands that lie above the Fermi level
at subcritical Hund coupling.  A similar prediction was made recently by
Bang on the basis of a phenomenological model for a heavily hole-doped
iron-pnictide superconductor\cite{bang_14}.

Now recall that the hydrodynamic Hamiltonian (\ref{h_hydro})
predicts a spin-$0$ collective $d$-wave oscillation
of the  $S^{+-}$ Cooper pair at long wavelength.
The $s/d$-wave energy-splitting density in a macroscopic superconductor
at a dilute number of hole pairs $N$
is $2 e_{J^{\prime}} \cong (N/V) (E_D - E_S)$.
Using the dependence $E_D - E_S \propto {\rm max}\,(0, J_0-J_{0c})$
for the $s/d$-wave energy splitting suggested by Fig. \ref{dE0vsJ0}b
yields the following expression for the dependence of
the excitation energy of the spin-$0$ collective mode: 
$\hbar\omega_{J^{\prime}}\propto (N/V)^{1/2} \cdot {\rm Re}\, (J_0-J_{0c})^{1/2}$.
It collapses to zero at the QCP and/or at half-filling.
This result also implies that $\hbar\omega_{J^{\prime}} = 0$
at super-critical Hund coupling,
where $\Delta_{cSDW} = 0$. 
The groundstate $S^{+-}$ Cooper pair shown in Fig. \ref{spctrm}, hence,
is stable {\it only} at subcritical Hund coupling.
More generally, the barrier for internal Josephson tunneling
$e_{J^\prime}$ vanishes at the QCP.
In such case, 
the relative phase $\phi_{-}$ winds freely,
which destroys internal phase coherence in 
the Cooper pair wavefunction (\ref{psi_phi}).
The quantum critical state\cite{qcp} at the QCP therefore shows
net superconducting phase coherence at long range,
which couples to electric charge,
but it shows no internal phase coherence
that can discriminate between
$s$, $d$, or $s+id$ pair symmetry,
for example, at long range.  [Cf. Eq. (\ref{s+id}).]
The previous exact results for two holes in
the two-orbital $t$-$J$ model (\ref{tJ}) confirm the 
collapse of the collective-mode spectrum to zero excitation energy
predicted above, but at short wavelength.
In particular,
the groundstate spin-$0$ energy levels at cSDW momenta shown in Fig. \ref{spctrm}
fall below the energy level of
the $s$-wave Cooper pair
as Hund coupling increases past the QCP.

\subsection{Local Cooper Pairs}
Figure \ref{dE0vsJ0} strongly suggests that the $d$-wave Cooper pair remains higher in
energy than the groundstate $s$-wave Cooper pair, despite the fact that the energy
splitting tends to zero at the QCP.  Does this ranking in energy persist in
the thermodynamic limit at a dilute concentration of charge carriers?
And does long-range phase coherence prevail in such case? 
We answer these two questions affirmatively below
on the basis of the local nature of the Cooper pairs.

The exact results shown in Fig. \ref{s+-d+-} for the $s$-wave and $d$-wave pair wave functions
suggest the component wavefunctions per orbital that are listed in Table \ref{table3}.
They are short range and extend at most
an iron-pnictide lattice constant $a^{\prime} = \sqrt{2} a$.
Following the results of the three-state model introduced in Appendix \ref{ppndx_3_pairs},
the $d_{yz}^{2(+-)}$ and $d_{xz}^{2(+-)}$ pairs primarily experience
a diagonal Hamiltonian matrix element $-E_b$ and 
an off-diagonal Hamiltonian matrix element $-{\cal T}_0$ at the QCP.
They account, respectively, for the binding 
and for the internal Josephson tunneling between the component pair states.
A positive sign for the tunneling parameter ${\cal T}_0$
yields an energy splitting between
an $s$-wave groundstate and a $d$-wave excited state.
Trial BCS wavefunctions for $S^{+-}$ and $D^{+-}$ pairs (\ref{psi_phi})
that project out double-occupancy at an iron site per $d\pm$ orbital
exhibit promising energies at half filling\cite{jpr_haas_15}.
Following Leggett's results for unprojected BCS states\cite{leggett_80,nozieres_schmitt-rink_85},
it plausible then to assume that the groundstate at
a dilute concentration of charge carriers
is a Bose condensate of local $S^{+-}$ Cooper pairs
that shows long-range phase coherence.

This picture of a Bose condensate of local Cooper pairs 
is encoded by the Ginzburg-Landau free-energy functional 
to lowest order,
\begin{equation}
{\cal F}_{cond}^{(2)} = |\bm{\nabla}\bm{\Psi}|^2/2 m_* - E_b |{\bm\Psi}|^2
-{\cal T}_0( \Psi_x \Psi_y^* + \Psi_x^* \Psi_y ),
\label{GL}
\end{equation}
where ${\bm\Psi} = (\Psi_x,\Psi_y)$ is the two-component order parameter
for $d_{yz}^{2(+-)}$ and $d_{xz}^{2(+-)}$ pairing, respectively.
Above, $m_*$ is the effective mass of a pair of holes.
The coherence length extracted from (\ref{GL})
is therefore
$\xi_0 = (2m_* E_b)^{-1/2}$.
Taking $E_b \cong 0.4\, J_1^{\perp}$ 
for the binding energy
from the spectrum displayed by Fig. \ref{spctrm}
and $m_*^{-1} = |t_1^{\parallel}| a^2$ for the inverse effective
mass yields the estimate
$\xi_0 \cong (|t_1^{\parallel}|/0.8 J_1^{\perp})^{1/2} a \cong 0.70$ nm
for the coherence length in iron-based superconductors.
Here, hopping coincides with Fig. \ref{FS}c,
and $a = 2.8\,{\rm \AA}$ is the intra-layer Fe-Fe distance
in optimally hole-doped (Ba,K)Fe$_2$As$_2$ \cite{rotter_prl_08}.
Experimental determinations of 
the upper-critical magnetic field along the $c$ axis
in the hole-doped series of iron-pnictide superconductors
Ba$_{1-x}$K$_x$Fe$_2$As$_2$
find steep slopes versus temperature at $T_c$
near optimal doping\cite{wang_prbrc_08,sun_prb_09}.
In particular, determinations of the irreversibility field along the $c$ axis
at $x\cong 0.4$ 
find that the product of this slope with $T_c$
is\cite{pissas_12} $H_1 = -T_c (d H_{c2}/d T)|_c = 2.5$ MOe.  
Substitution into the formula
$\xi_1 = (\Phi_0/2\pi H_1)^{1/2}$
for the Ginzburg-Landau coherence length,
with the magnetic flux quantum
$\Phi_0 = 2.07 \times 10^{-7}$ G cm$^2$,
yields the estimate $\xi_1 = 1.16$ nm.
It is comparable to the previous estimate for the coherence length
based on local Cooper pairs.

\subsection{Dispersion of Two-Hole Spectrum}
The nature of the low-energy spectrum
displayed by Fig. \ref{spctrm}
for a pair of holes
roaming over a $4\times 4$ lattice of spin-$1$ iron atoms (\ref{tJ})
with {\it net} 2D momentum
further corroborates the identification of the groundstate at zero 2D momentum
as a $S^{+-}$ Cooper pair.  In particular,
the groundstate at cSDW wavenumber $(\pi/a){\bm{\hat x}}$ is spin-$0$,
and it is even and odd under reflections about
the $x$ axis and the $y$ axis, respectively.  It is therefore consistent with 
a moving $|d_{yz}^{2(+-)}\rangle$ Cooper pair state:
${\rm sin}[(\pi/2a) (x_1 + x_2)] \langle 1, 2|d_{yz}^{2(+-)}\rangle$.
Indeed,
a direct computation of the pair amplitude (\ref{iF})
finds that it is proportional to
${\rm sin}\, k_x a$ at $k_y = 0$
in the  $d_{yz}$-$d_{yz}$ pair channel, 
while it is approximately zero otherwise.
(Cf. Table \ref{table3}.)
Next, the lowest-energy spin-$1$ state
in Fig. \ref{spctrm} at this cSDW momentum 
is the critical cSDW spin resonance,
which can be interpreted as a pair excitation of
the $S^{+-}$ groundstate\cite{korshunov_eremin_08,maier_scalapino_08}.
It has even parity under a reflection about the $x$ axis, as well as about
the $y$ axis.  
The spin resonance at wavenumber $(\pi/a){\bm{\hat x}}$ 
can therefore also be interpreted as a moving $P_x$ triplet Cooper pair:
${\rm sin}[(\pi/2a) (x_1 + x_2)] \langle 1, 2|P_x\rangle$.
Also, both the first-excited spin-$0$ and spin-$1$ states
at wavenumber $(\pi/a){\bm{\hat x}}$
are odd and even
under reflections about the $x$ axis and the $y$ axis, respectively.
They may therefore be interpreted as particle-hole excitations of
the groundstate $S^{+-}$ Cooper pair.
Further,
the spin-$0$ and spin-$1$ groundstates at 
wavenumbers $(1,1)$ and $(2,2)$ in units of $\pi/2a$
share the same parity under a true reflection about the $x$-$y$ diagonal,
which includes swap of the $d_{xz}$ and $d_{yz}$ orbitals.  
In particular, the even and odd parity states are degenerate per spin.
We therefore also interpret these low-lying states as particle-hole excitations
of the groundstate $S^{+-}$ Cooper pair.
Finally, both the spin-$0$ and spin-$1$ groundstates at momentum $(2,1)$ 
in Fig. \ref{spctrm} are even under a reflection about the $y$ axis.
The weak dispersion in energy per spin at nearby momenta suggests that the former
is a spin-$0$ collective excitation of the groundstate $S^{+-}$ Cooper pair
and that the latter is a spin resonance of the same.


\section{Summary and Conclusions}
We have found
groundstate $S^{+-}$ and excited-state $D^{+-}$ Cooper pairs
near a QCP where they become degenerate
in a local-moment model for iron-based superconductors.
Such Cooper pairing instabilities were predicted previously
for iron-pnictide superconductors on the basis of
perturbative spin-fluctuation exchange\cite{wang_09}.
In the present local-moment model, however,
the $S^{+-}$ and $D^{+-}$ Cooper pairs
are respectively bonding and anti-bonding superpositions of orbitally ordered Cooper pairs
that alternate in sign between hole and electron Fermi surfaces.
With the exception of $d_{xy}$ orbital character 
at the tips of the electron pockets\cite{sup_mat_3},
this result is consistent with
the $s$-wave and $d$-wave solutions to the gap equation 
reported by Graser et al. in ref. \cite{graser_09},
which again invokes spin-fluctuation exchange,
but within the random-phase approximation.
Note that
the binding energy of the $s$-wave singlet pair groundstate 
at the QCP shown in Fig. \ref{spctrm}
implies a gap 
$E_b = 0.4\, J_1^{\perp}$.
Also,
the inset to Fig. \ref{dE0vsJ0}a
indicates that this quasi-particle gap does {\it not} collapse to zero at the QCP.
Taking a value of $J_1^{\perp}\sim 120$ meV from 
a fit to spin-wave spectra in iron-pnictide materials 
based on the corresponding two-orbital
Heisenberg model (Fig. \ref{FS}a and ref. \cite{jpr_10})
then yields a gap $E_b \sim 50$ meV.
It is roughly consistent with the upper bound for the spin-resonance energy
in iron-pnictide superconductors\cite{paglione_greene_10},
and it is consistent with high-temperature superconductivity.

Our focus on the degenerate $d_{xz}$ and $d_{yz}$ orbitals 
that feature prominently in iron superconductors
reveals that the $S^{+-}$ and $D^{+-}$
bound pair states are linked by a collective excitation
within a larger family of $s+id$ pair states (\ref{psi_phi}).
Observe, in particular, that such pair wavefunctions
can be re-expressed as
\begin{equation}
|\psi_{J^{\prime}}(\phi_-)\rangle = ({\rm cos}\, {1\over 2}\phi_-)|S^{+-}\rangle
+ i ({\rm sin}\, {1\over 2}\phi_-)|D^{+-}\rangle.
\label{s+id}
\end{equation}
The stable $s$-wave state notably passes through
the $s+id$ state on route to the unstable $d$-wave state\cite{khodas_chubukov_12}.
Previously,   
Scalapino and Devereaux
predicted a related $d$-wave exciton
for the $S^{+-}$ state at weak coupling\cite{s&d_09}, 
with the important distinction that
they required nested electron-type Fermi surface pockets at cSDW momenta.
Recent studies of Raman spectroscopy
in the optimally hole-doped iron-pnictide superconductor 
Ba$_{0.6}$K$_{0.4}$Fe$_2$As$_2$
report experimental evidence for an in-gap collective mode\cite{raman_13,raman_14}
consistent with the $s$-to-$d$-wave pair excitation
identified here and in ref. \cite{s&d_09}.
ARPES demonstrates that the electron pockets at cSDW mometa lie below the Fermi level
at optimal doping\cite{ding_08,ding_09,nakayama_prb_11}.
Within the present theory,
optimal doping corresponds to super-critical Hund coupling,
at which the frequency $\omega_{J^{\prime}}$ of the collective mode
has collapsed to a small value.
The present theory predicts, however, that the collective mode
persists at the Lifshitz transition, 
where the electron Fermi surface pockets vanish,
and that the collective oscillation frequency has a similarly small value there.
On this basis, we therefore predict that the $s$-to-$d$-wave collective mode
observed in Ba$_{1-x}$K$_{x}$Fe$_2$As$_2$ at optimal doping\cite{raman_13,raman_14}
will also be observed at higher doping 
near the Lifshitz transition\cite{malaeb_prb_12,budko_prb_13,hodovanets_prb_14}.

\begin{acknowledgments}
The author thanks Xing-Jiang Zhou, Peter Hirschfeld, Thomas Maier,
Paolo Zanardi and Stephan Haas for valuable discussions.
He also thanks Brent Andersen, Richard Roberts and Timothy Sell for 
technical help with the use of  
the virtual shared-memory cluster (Lancer) at
the AFRL DoD Supercomputing Resource Center.
This work was completed in part during the 2014 program
on ``Magnetism, Bad Metals and Superconductivity: Iron Pnictides and Beyond''
held at the Kavli Institute for Theoretical Physics,
with partial support from the National Science Foundation (NSF)
under grant no. PHY11-25915.
This work was also supported in part by the US Air Force
Office of Scientific Research under grant no. FA9550-13-1-0118
and by the NSF PREM program under grant no. DMR-1523588.
\end{acknowledgments}

\appendix
\section{{\label{ppndx_spin_gap}}Quantum Critical Point at Half Filling}
Near cSDW wavenumbers
${\bm Q}_{cSDW} = (\pi/a){\bm{\hat x}}$ and $(\pi/a){\bm{\hat y}}$,
and in the absence of mobile holes,
the spin-wave spectrum
for the two-orbital $t$-$J$ model (\ref{tJ})
disperses anisotropically as
\begin{equation}
\omega({\bm k}) =
[\Delta_{cSDW}^2 + v_{l}^2 (k_{l} - \pi / a)^2 + v_{t}^2 k_{t}^2]^{1/2}
\label{csdw_dispersion}
\end{equation}
within the semi-classical approximation
about hidden magnetic order,
$\nwarrow_{d-}\searrow_{d+}$,
at large electron spin\cite{jpr_10}, $s_0\rightarrow\infty$.
Here, $k_l$ and $k_t$ denote the components of wavenumber ${\bm k}$
that are parallel and perpendicular to ${\bm Q}_{cSDW}$.
Above, the spin gap at ${\bm Q}_{cSDW}$ collapses to zero at the QCP as
\begin{equation}
\Delta_{cSDW} = (2 s_0) [(4 J_2^{\perp} - J_{0c}) (J_0 - J_{0c})]^{1/2},
\label{delta_csdw}
\end{equation}
while the longitudinal  spin-wave velocity $v_l$ and 
the anisotropy parameter $v_l/v_t$ coincide with the values
\begin{equation}
v_0 = 2 s_0 a  [(J_1^{\perp} - J_1^{\parallel} + 2 J_2^{\perp} - 2 J_2^{\parallel})
\cdot ({1\over 2} J_0 + 2 J_1^{\perp} + 2 J_2^{\perp})]^{1/2}
\label{v0}
\end{equation}
and
\begin{equation}
\gamma_0 = 
\Biggl({2 J_2^{\parallel} + 2 J_2^{\perp} + J_1^{\parallel} + J_1^{\perp}
\over{2 J_2^{\parallel} + 2 J_2^{\perp} - J_1^{\parallel} - J_1^{\perp}}}\Biggr)^{1/2}
\label{anis_csdw}
\end{equation}
at criticality.
The hidden-order phase is stable at weak Hund coupling, $-J_0 < -J_{0c}$,
with
$-J_{0c} = 2 (J_1^{\perp} - J_1^{\parallel}) - 4 J_2^{\parallel}$.
The correlation length for cSDW order,
$\xi_{cSDW} = v_l/\Delta_{cSDW}$,
therefore diverges as $(J_0 - J_{0c})^{-1/2}$
at the QCP within the linear spin-wave approximation\cite{jpr_10}.
It implies a second-order quantum phase transition\cite{qcp},
with critical exponents $\nu_{cSDW} = 1/2$ and $z_{cSDW} = 1$.
Exact results for the same two-orbital Heisenberg model
over a $4\times 4$ lattice of iron atoms
yield that
the square of the  hidden-order moment and the square of the cSDW-order moment
dove-tail at the QCP\cite{jpr_10}.
This also suggests a second-order 
quantum phase transition at the QCP in the thermodynamic limit.

At the long-wavelength limit,
the spin-wave spectrum follows
$\omega({\bm k}) = v_0 |{\bm k}|$ in the hidden-order phase
within the semi-classical approximation. 
It has no spectral weight in the true spin channel, however.
(See Figs. \ref{FS}a,b.)

\section{{\label{ppndx_3_pairs}}Three-State Model}
The lowest-energy pair states in the two-orbital $t$-$J$ model at the QCP 
shown by the spectrum in Fig. \ref{spctrm}
can be described by the following effective Hamiltonian
in the reduced Hilbert space composed of the pair states
$|x\rangle = |d_{yz}^{2(+-)}\rangle$,
$|y\rangle = |d_{xz}^{2(+-)}\rangle$, and
$|z\rangle = |S^{++}\rangle$:
\begin{equation}
H_{pair} =
\begin{pmatrix}
-E_{b}   & -{\cal T}_0 & -{\cal T}_1 \\
-{\cal T}_0 & -E_{b}      & -{\cal T}_1 \\
-{\cal T}_1 & -{\cal T}_1 & E_z \\
\end{pmatrix} .
\label{3pair}
\end{equation}
Above, $E_b > 0$ is the binding energy of the 
degenerate pair states $d_{yz}^{2(+-)}$ and $d_{xz}^{2(+-)}$,
 $E_z > 0$ is the energy of the $S^{++}$ Cooper pair,
and   ${\cal T}_0$ and ${\cal T}_1$ are
positive and real hybridization matrix elements,
respectively,
among the three pair states.
Zero energy is set by the edge of the continuum.
The eigenstates of (\ref{3pair}) can be easily found.
They have the form
\begin{eqnarray}
|0\rangle &=& 
({\rm cos}\,\theta_0) |S^{+-}\rangle + ({\rm sin}\,\theta_0) |S^{++}\rangle, \nonumber \\
|1\rangle &=&
|D^{+-}\rangle, \nonumber \\
|2\rangle &=& 
-({\rm sin}\,\theta_0)|S^{+-}\rangle + ({\rm cos}\,\theta_0) |S^{++}\rangle,
\label{3states}
\end{eqnarray}
where above we call
$|S^{+-}\rangle = {1\over{\sqrt 2}}|d_{yz}^{2(+-)}\rangle + {1\over{\sqrt 2}}|d_{xz}^{2(+-)}\rangle$
and
$|D^{+-}\rangle = {1\over{\sqrt 2}}|d_{yz}^{2(+-)}\rangle - {1\over{\sqrt 2}}|d_{xz}^{2(+-)}\rangle$.
Application of $H_{pair}$ yields energy eigenvalues of the form
\begin{eqnarray}
E_0 &=& -E_{b} -{\cal T}_0 - \epsilon_0, \nonumber \\
E_1 &=& -E_{b}+{\cal T}_0, \nonumber \\
E_2 &=&  E_z + \epsilon_0.
\label{3energies}
\end{eqnarray}
Above, $\theta_0$ is set by
\begin{equation}
{\rm tan}\,\theta_0 = 
({\rm sgn}\,{\cal T}_1)\sqrt{\Biggl({{\cal T}_0+\delta E\over{2^{3/2} {{\cal T}_1}}}\Biggr)^2 + 1}
-{{\cal T}_0+\delta E\over{2^{3/2} {{\cal T}_1}}},
\label{tan_theta_0}
\end{equation}
where $\delta E = E_z + E_b$
is the energy difference between the $S^{++}$ pair state
and the degenerate Cooper pairs $d_{yz}^{2(+-)}$ and $d_{xz}^{2(+-)}$.
Also, $\epsilon_0 = 2^{1/2} {\cal T}_1\, {\rm tan}\,\theta_0$
is a level-repulsion energy.

We shall now  assume 
that both ${\cal T}_0$ and ${\cal T}_1$ tend to zero at the QCP.
This yields the approximations
\begin{eqnarray}
\theta_0 &\cong& 2^{1/2} {\cal T}_1/\delta E, \\
\epsilon_0 &\cong& 2 {\cal T}_1^2/\delta E
\label{theta_epsilon_qcp}
\end{eqnarray}
for the mixing angle and for the energy-level repulsion near the QCP,
at ${\cal T}_0,{\cal T}_1 \ll \delta E$.
Inspection of (\ref{3states}) and (\ref{3energies}) yields that they
are negligibly small in that critical region.

%
%

%
%

\pagebreak
\widetext
\begin{center}
\textbf{\large Supplemental Material: Collective Mode at Lifshitz Transition in Iron-Pnictide Superconductors}
\end{center}

\bigskip
\begin{center}
\text{Jose P. Rodriguez}
\end{center}

\medskip
\begin{center}
\it{Department of Physics and Astronomy,}
\end{center}
\begin{center}
\it{California State University at Los Angeles, Los Angeles, California 90032.}
\end{center}

\setcounter{equation}{0}
\setcounter{figure}{0}
\setcounter{table}{0}
\setcounter{page}{1}
\makeatletter
\renewcommand{\theequation}{S\arabic{equation}}
\renewcommand{\thefigure}{S\arabic{figure}}
\renewcommand{\bibnumfmt}[1]{[S#1]}
\renewcommand{\citenumfont}[1]{S#1}


\bigskip\bigskip
\begin{center}
\textbf{I. Mott Transition versus Structural Transition}
\end{center}
\bigskip

\begin{figure}
\includegraphics[scale=0.70, angle=-90]{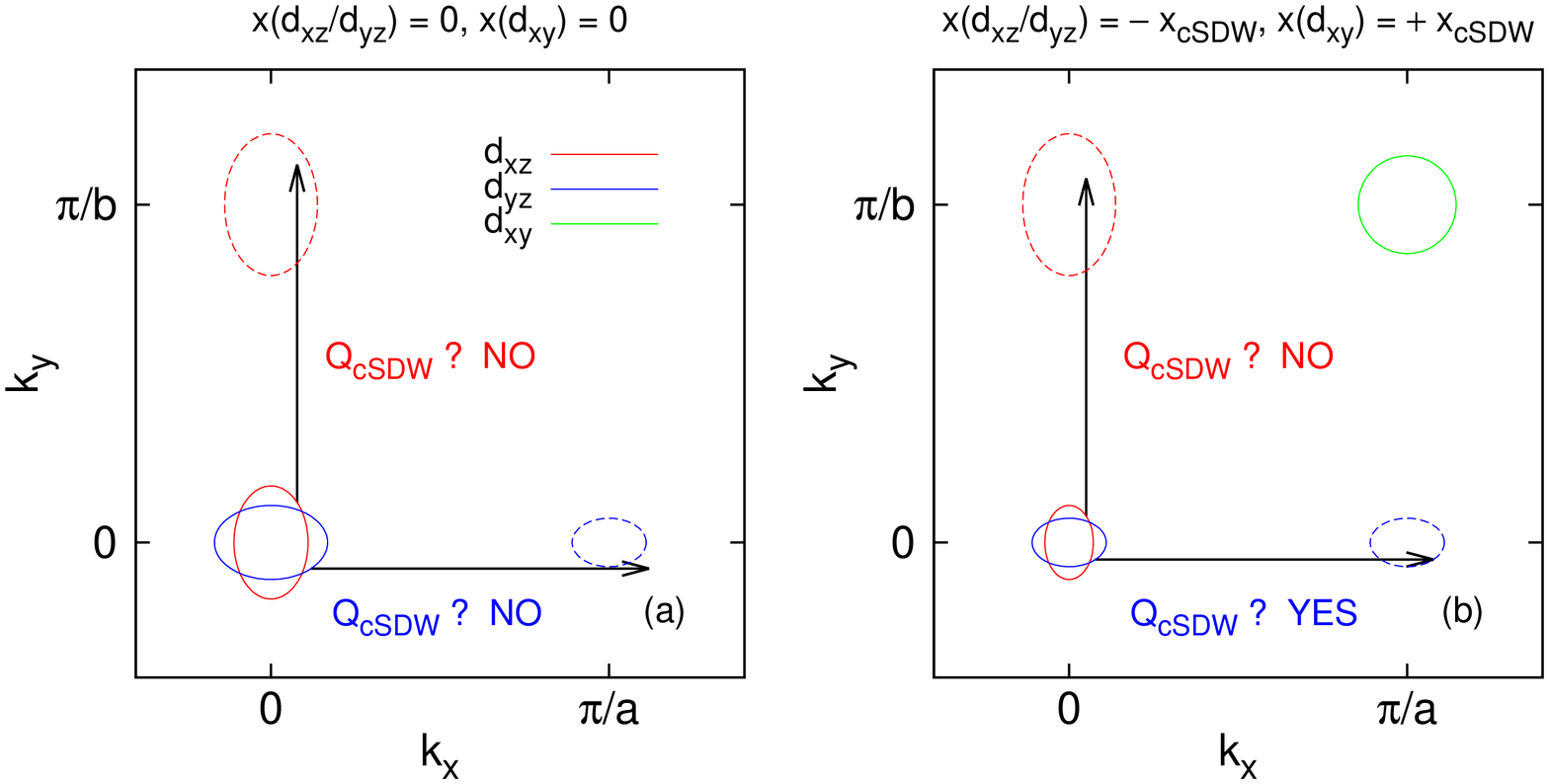}
\caption{Fermi surfaces of the two-orbital $t$-$J$ model
(a) at and (b) near half filling.
The emergent electron bands 
in the $d_{yz}$ and in the $d_{xz}$ orbitals
are shifted up and down in energy
because of an applied orthorhombic shear strain,
$(a-b)/(a+b) > 0$.
(Cf. ref. \cite{s_yoshizawa_simayi_12}.)
The labels $k_x$ and $k_y$ denote components of pseudo momentum.
(See ref. \cite{s_Lee_Wen_08}).}
\label{FS+strain}
\end{figure}

\begin{figure}
\includegraphics[scale=0.70, angle=-90]{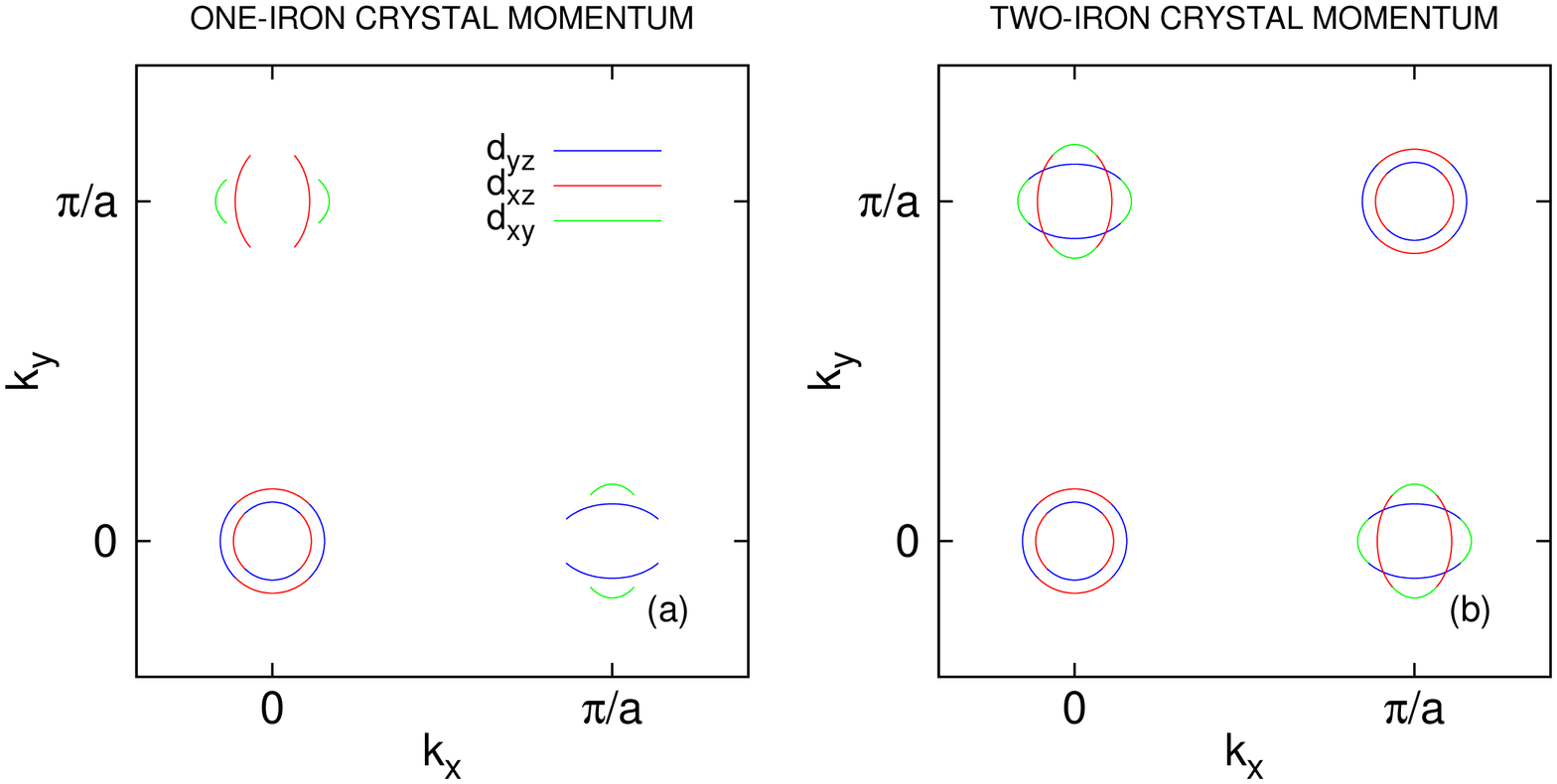}
\caption{Sketch of the typical Fermi surfaces predicted 
by DFT for iron-pnictide materials
near half filling. (See ref. \cite{s_kemper_10}.) 
Panels (a) and (b) correspond,
respectively, to equivalent and to inequivalent iron atoms
per iron-pnictide unit cell.}
\label{DFT_FS}
\end{figure}

The two-orbital $t$-$J$ model [Eq. (1) in the paper]
predicts a Mott insulator groundstate at half-filling.
Parent compounds to iron-pnictide superconductors at half-filling 
are bad metals\cite{s_nakajima_14},
on the other hand,
with orthorhombic crystal structure at low temperature\cite{s_nandi_10}.  
Let us simulate the last fact 
by imposing an orthorhombic shear strain 
\begin{equation}
\epsilon_{x^{\prime}y^{\prime}} = 
{1\over 2} \Biggl({\partial u_x \over{\partial x}} - {\partial u_y\over{\partial y}}\Biggr)
= {a-b\over{a+b}} ,
\label{shear_strain}
\end{equation}
with new lattice constants $a > b$
such that $a_0^2 = a b$.
Here $a_0$ is the original tetragonal lattice constant.
Assume now that the shear strain couples to the emergent electron bands so that
the $d_{yz}$ band at $(\pi/a_0){\bm{\hat x}}$ and
the $d_{xz}$ band at $(\pi/a_0){\bm{\hat y}}$
respectively 
shift rigidly up in energy and
shift rigidly down in energy
by an equal amount
proportional to $\epsilon_{x^{\prime}y^{\prime}}$.
(Cf. ref. \cite{s_yoshizawa_simayi_12}.)
Such a nematic asymmetry between the electronic orbitals may be intrinsically
due to the orthorhombic strain,
or vice versa\cite{s_nandi_10}.
In particular,
the orthorhombic crystal structure may induce anisotropy
in the Heisenberg exchange coupling constants so that 
commensurate spin-density wave (cSDW) order is
established preferentially along the $a$ axis\cite{s_schmidt_10},
or nematic symmetry breaking could occur
because of antiferromagnetic frustration\cite{s_xu_muller_sachdev_08}.
Figure \ref{FS+strain}a depicts 
the resulting emergent Fermi surfaces\cite{s_jpr_mana_pds_14}
of the two-orbital $t$-$J$ model at the QCP,
in the Mott insulator state at half filling.
For the sake of clarity,
we have turned off hybridization between 
hole bands with $d_{xz}$ and $d_{yz}$ orbital character.
(Cf. Fig. \ref{DFT_FS}.)
The Fermi surfaces are no longer nested
by $(\pi/a)\bm{\hat x}$ or by $(\pi/b)\bm{\hat y}$.
Notice also that the system is now orbitally ordered\cite{s_ono_10},
with more electrons populating the $d_{xz}$ versus the $d_{yz}$ orbital.

Emergent nesting along the $a$ axis can be restored if
electrons from other orbitals, such as the $d_{xy}$ one,
migrate into the $d_{xz}/d_{yz}$ hole bands. 
Figure \ref{FS+strain}b depicts 
the new nesting of the Fermi surfaces in such case.
It bears some resemblance to
the nematicity in the electronic structure of orthorhombic/cSDW parent compounds
to iron-pnictide superconductors revealed experimentally by
angle-resolved photoemission spectroscopy (ARPES)\cite{s_yi_11}.
Notice the appearance of a new $d_{xy}$ hole pocket that compensates for
the reduced area of the $d_{xz}/d_{yz}$ hole pockets
in Fig. \ref{FS+strain}b.
Density-functional theory (DFT) calculations
predict that such a $d_{xy}$ hole pocket appears 
upon hole doping from half filling
in the absence of electronic nematicity\cite{s_kemper_10}.
The presence of the $d_{xy}$ hole pocket at half-filling
shown in Fig. \ref{FS+strain}b
could be driven by the gain in magnetic energy 
that results from establishing cSDW order along the $a$ axis.
In such case, however, the system is no longer a Mott insulator.
Instead, it is a bad metal at half filling,
with a low concentration of electrons
in the $d_{xz}/d_{yz}$ orbitals overall, 
$x_{cSDW}\propto \epsilon_{x^{\prime}y^{\prime}}$,
compensated by an equal concentration of holes
in the $d_{xy}$ orbital.

\bigskip\bigskip
\begin{center}
\textbf{II. Electron $d_{xy}$ Pairing at Strong On-site Coulomb Repulsion}
\end{center}
\bigskip

The limit of strong on-site Coulomb repulsion,
which is adopted in the paper,
puts severe restrictions
on the pairing symmetry in iron-pnictide superconductors.  
Let us define the pairing function of the superconducting groundstate in the usual way:
$iF_{i,\alpha;j,\beta} =
\langle 
{c}_{i, \alpha,\uparrow}^{\dagger}
{c}_{j, \beta,\downarrow}^{\dagger}\rangle$,
where $i$ and $j$ denote iron sites,
and where $\alpha$ and $\beta$ are $d_{xz}$, $d_{yz}$ or $d_{xy}$ orbitals.
The limit $U_0, U_0^{\prime}\rightarrow\infty$ then imposes the constraints
$iF_{i,\alpha;i,\beta} = 0$ at all sites $i$.
Application of translational invariance yields one constraint,
\begin{equation}
\sum_{\bm k} \langle
{c}_{\alpha,\uparrow}({\bm k})^{\dagger}
{c}_{\beta,\downarrow}(-{\bm k})^{\dagger}\rangle = 0 .
\label{pair_constraint}
\end{equation}
The sum in momentum above is over the one-iron (unfolded) Brillouin zone,
while
$c_{\alpha,s}({\bm k}) = N_{\rm Fe}^{-1/2} \sum_i
e^{-i {\bm k}\cdot{\bm r}_i} c_{i,\alpha,s}$
destroys a spin-$s$ electron in orbital $\alpha$ that carries momentum $\hbar \bm k$.
The latter coincides with the crystal momentum 
if the two iron sites per iron-pnictide unit cell are equivalent.
It represents the pseudo momentum connected with 
the glide-reflection symmetry shown by 
an isolated iron-pnictide layer
if the two iron atoms per unit cell are inequivalent\cite{s_Lee_Wen_08}.
The pseudo momentum coincides with the one-iron crystal momentum
in the case of an electron in a $d_{xz}/d_{yz}$ orbital,
while these two momenta differ by a
reciprocal lattice vector $(\pi/a)({\bm{\hat x}} \pm {\bm{\hat y}})$
in the case of an electron in the $d_{xy}$ orbital.
Figure \ref{DFT_FS} depicts a typical Fermi surface for iron-pnictide materials
in the one-iron Brillouin zone and in the two-iron Brillouin zone
as predicted by density-functional theory\cite{s_kemper_10}.
Below, we list the restrictions on $d_{xy}$ pairing
when constraint (\ref{pair_constraint}) is enforced.

{\it S-wave Cooper Pair.}
Suppose that $\alpha = d_{xy} = \beta$ in the constraint (\ref{pair_constraint}).
Then study of the Fermi surfaces depicted in Figs. \ref{DFT_FS}a and \ref{DFT_FS}b
imply that isotropic pairing is ruled out.
Next, suppose instead that $\alpha = d_{xy}$ and that $\beta = d_{yz}$.
Again, the constraint (\ref{pair_constraint}) is incompatible 
with Figs. \ref{DFT_FS}a and \ref{DFT_FS}b,
which means that $s$-wave pairing is ruled out once more.

{\it D-wave Cooper Pair.}
Consider again the diagonal orbital channel $\alpha = d_{xy} = \beta$.
Figure \ref{DFT_FS} shows that $D_{x^2-y^2}$ pairing is possible in this channel,
which satisfies the constraint (\ref{pair_constraint}).
In the particular case of inequivalent iron atoms,
Fig. \ref{DFT_FS}b indicates that $d_{xz}/d_{yz}$ pairing is nodeless,
while that $d_{xy}$ pairing has nodes.  This implies that the
former is the dominant pairing channel in such case.
In the off-diagonal orbital channel,
$\alpha = d_{xy}$ and $\beta = d_{yz}$,
Fig. \ref{DFT_FS}a in conjunction with (\ref{pair_constraint}) imply
that $D_{x^2-y^2}$ pairing is ruled out in the case of equivalent iron atoms.
The same holds true
in the case of inequivalent iron sites, Fig. \ref{DFT_FS}b.
In the latter case, specifically,
$D_{xy}$ pairing is possible in this off-diagonal orbital channel, 
but it has nodes.
It will therefore be smaller in amplitude than the dominant $d_{xz}/d_{yz}$
orbital channel for $D_{x^2-y^2}$ pairing discussed previously.

{\it P-wave Cooper Pair.}
Study of Fig. \ref{DFT_FS} yields that $P_x$ triplet pairing is generally possible
in the diagonal orbital channel $\alpha = d_{xy} = \beta$,
which again satisfies the constraint (\ref{pair_constraint}).
Both this pair wave function 
and the $d_{yz}$ counterpart show nodes.
This indicates that they are competing pair states
with comparable magnitudes.
Next, set $\alpha = d_{xy}$ and $\beta = d_{xz}$
in the constraint (\ref{pair_constraint}).
Inspection of Fig. \ref{DFT_FS} yields that $P_x$ triplet pairing is possible
in this off-diagonal orbital channel as well,
but only in the case of inequivalent iron sites.
In principle,
it is comparable in magnitude to
$d_{yz}$-$d_{yz}$ triplet $P_x$ pairing,
but it lives in a much reduced momentum space in between the principal axes.


\begin{thebibliography}{}
\bibitem{heuristic_pBCS} An alternative heuristic form for 
the condensate of $2N$ holes is given by the ``binomial'' expression
%
\begin{equation}
\langle 1, ..., 2N|\Psi_N\rangle =
{1\over{\sqrt{2^N (2N)!}}}
\sum_p ({\rm sgn}\, p) \langle p(1),p(2)|\psi_{J^{\prime}}\rangle 
... \langle p(2N-1),p(2N)|\psi_{J^{\prime}}\rangle , \nonumber
\end{equation}
%
where $p$ are permutations.
\end{thebibliography}
\end{document}